\documentclass[11pt]{article}
\usepackage{jheppubmod}
\usepackage{color}
\usepackage[showonlyrefs]{mathtools}
\usepackage{psfrag}
\usepackage{array}
\usepackage{amssymb}
\usepackage{graphicx}
\usepackage{amsmath}
\usepackage{empheq}
\usepackage{slashed}
\usepackage{caption}
\usepackage[labelsep=quad]{subcaption}
\usepackage{epstopdf}
	
\usepackage{epsfig}
\usepackage[punctsep]{collref}
\newcommand{\be}{\begin{equation}}
\newcommand{\ee}{\end{equation}}
\collectsep[]{;}	
\usepackage{cite}
 
\allowdisplaybreaks

\def\LL{{\cal L}}

\def\NN{{\cal N}}
\def\OO{{\cal O}}

\def\VV{{\cal V}}

\def\vac{|0\rangle}

\def\q{{\bf q}}

\def\beq{\begin{equation}}
\def\eeq{\end{equation}}
\newcommand{\bea}{\begin{eqnarray}}
\newcommand{\eea}{\end{eqnarray}}

\title{
On the K\"ahler-Hodge structure of superconformal manifolds
}

\author[a]{Vasilis Niarchos,}
\author[b]{Kyriakos Papadodimas}
\affiliation[a]{CCTP and ITCP,
Department of Physics, University of Crete, 71303, Greece}
\affiliation[b]{Theoretical Physics Department, CERN, CH-1211 Geneva 23, Switzerland}
\emailAdd{niarchos@physics.uoc.gr}
\emailAdd{kyriakos.papadodimas@cern.ch}

\preprint{ITCP-IPP 2021/5\\ 
\hspace*{0.1cm} \hfill CCTP-2021-9\\
\hspace*{0.1cm} \hfill CERN-PH-TH-2021-226\\}

\date{}

\abstract{
We show that conformal manifolds in $d\geq 3$ conformal field theories with at least 4 supercharges are K\"ahler-Hodge, thus extending to 3d ${\cal N}=2$ and 4d ${\cal N}=1$ similar results previously derived for 4d ${\cal N}=2$ and ${\cal N}=4$ and various types of 2d SCFTs. Conformal manifolds in SCFTs are equipped with a holomorphic line bundle ${\cal L}$, which encodes the operator mixing of supercharges under marginal deformations. Using conformal perturbation theory and superconformal Ward identities, we compute the curvature of ${\cal L}$ at a generic point on the conformal manifold. We show that the K\"ahler form of the Zamolodchikov metric is proportional to the first Chern class of ${\cal L}$, with a constant of proportionality given by the two-point function coefficient of the stress tensor, $C_T$.  In cases where certain additional conditions about the nature of singular points on the conformal manifold hold, this implies a quantization condition for the total volume of the conformal manifold.}

\keywords{Supersymmetry, superconformal field theories}

\begin{document}
\maketitle

\section{Introduction}

In this paper we show that conformal manifolds in $d\geq 3$ superconformal field theories (SCFTs) with at least 4 supercharges are K\"ahler-Hodge. K\"ahler-Hodge manifolds are K\"ahler manifolds with the property that the flux of the K\"ahler form through any two-cycle is quantized. This extends to 4d ${\cal N}=1$ and 3d ${\cal N}=2$ theories similar results previously found for 4d ${\cal N}=2,4$ theories \cite{Papadodimas:2009eu,Gomis:2015yaa}, as well as 2d ${\cal N}=(2,2)$ SCFTs (see for instance \cite{Strominger:1990pd,Bershadsky:1993cx}). 

A central element in our discussion is the existence of a line bundle ${\cal L}$ over the conformal manifold ${\cal M}$ of superconformal field theories, which encodes the operator mixing between supercharges under conformal perturbation theory. The superconformal algebra is invariant under an automorphism which, for example in the case of 4d ${\cal N}=1$ theories, rotates the supercharges and superconformal partners as
\be
\begin{split}
& Q_a \rightarrow e^{i\theta} Q_a \qquad , \qquad \overline{Q}_{\dot{a}} \rightarrow e^{-i\theta} \overline{Q}_{\dot{a}}\cr
& S^a \rightarrow e^{-i\theta} S^a \qquad , \qquad \overline{S}^{\dot{a}} \rightarrow e^{i\theta} \overline{S}^{\dot{a}}
\end{split}
\ee
leaving all other generators invariant. 
This automorphism implies that there is an in-principle coupling constant dependent ambiguity in the choice of this phase and that the supercharges should be thought of as being associated to sections of a line bundle ${\cal L}$ over the conformal manifold. We show that conformal perturbation theory naturally defines a non-trivial connection on this bundle. This can be interpreted as standard Berry phase when thinking of states on ${\mathbb S}^{d-1} \times {\rm time}$ using the state-operator map \cite{Baggio:2017aww}. 

As a preliminary result we show that the superconformal manifolds under investigation are K\"ahler. This generalizes previous results in the literature \cite{Asnin:2009xx}.

We then show that ${\cal L}$ is a holomorphic line bundle whose curvature is proportional to the K\"ahler form of the Zamolodchikov metric\footnote{The precise numerical coefficients depend on the conventions used in the definition of the Zamolodchikov metric, as explained in the main text.}
\be
\label{mainresultofpaper}
\begin{split}
&F_{ij}= F_{\overline{i}\,\overline{j}} = 0\cr 
F_{i\overline{j}}=-{1\over 768c} G_{i\overline{j}} &\qquad {\rm for}\qquad 4d\,\, {\cal N}=1,2,4\cr
F_{i\overline{j}}=-{1\over 4 \pi^2 C_T } G_{i\overline{j}} & \qquad {\rm for}\qquad 3d\,\, {\cal N}=2
\end{split}
\ee
In the above equations, $c$ is the conformal anomaly in 4d and $C_T$ is the 2-point function coefficient of the stress tensor in 3d ---see \eqref{conventionst} for the precise conventions.\footnote{We could write the 4d formula also in terms of $C_T$, using the relation $c= {\pi^4 \over 40} C_T$ in 4d (see e.g.\ \cite{Osborn:1993cr,Barnes:2005bm}).} 

This means that superconformal manifolds are not only K\"ahler, but moreover K\"ahler-Hodge, since the K\"ahler form $k$ associated to the Zamolodchikov metric is proportional to the first Chern class of a line bundle over ${\cal M}$, i.e.
\be
\label{kahlerm}
k=\alpha\, C_T F
\ee
where the numerical coefficient $\alpha$ is theory-independent, though it depends on whether we are in 3d or 4d and the precise conventions. It can be extracted from the equations \eqref{mainresultofpaper}.

The result \eqref{mainresultofpaper} may be useful in studying global properties of superconformal manifolds. For instance, it implies interesting quantization properties of the Zamoldochikov metric which we now discuss.

If an $n$-complex dimensional superconformal manifold is compact and has no singularities, then the fact that the K\"ahler form  is proportional to the first Chern class of a line bundle \eqref{kahlerm} implies that the total volume of the Zamoldchikov metric must be quantized in units of $C_T$
\footnote{
The quantization condition extends to the quantization of the K\"ahler flux (and hence to minimal volumes) over any compact even cycle without any singularities.}
\be
V_{\cal M} = {1\over n!}\int_{\cal M} k \wedge ...\wedge k = 
{(2\pi)^n \over n!}\alpha^n (C_T)^n \times{\rm integer}
~.
\ee
where the integer is determined by the Chern number of ${\cal L}$.
We emphasize again that the numerical constant $\alpha$, given in Eqs.\ \eqref{mainresultofpaper}, is theory-independent.

Known conformal manifolds in 4d ${\cal N}=2$ theories are usually non-compact, so more care has to be taken in the computation of the total volume (see, however, the discussion below). Compact conformal manifolds have been discussed in \cite{Buican:2014sfa,Baggio:2017mas}. In fact, Ref.\ \cite{Perlmutter:2020buo} has proposed that all 3d conformal manifolds may be compact. In some of these cases, the conformal manifolds have orbifold singularities, where again special analysis would be needed to understand the behavior of ${\cal L}$ near the singularities and how the latter could contribute to the total K\"ahler flux and hence volume.

If a superconformal manifold has only singularities of weak-coupling type, the previous statement can be extended as follows. Suppose we have a submanifold $P$, where the theory develops a weak-coupling singularity. We assume that near $P$ we can use local coordinates involving a complexified gauge coupling $\tau$,
\be
ds^2 = {\rm const} {d\tau d\overline{\tau} \over {\rm Im}\tau^2} +\ldots
~,
\ee
that tends to infinity ($\tau\to i \infty$) in the weak coupling limit, and additional exactly marginal couplings at regular, finite values. While the weakly coupled region $P$ is at infinite Zamolodchikov distance, the total volume of the hyperbolic cusp is finite. Moreover, one can check by a weak-coupling computation that the line bundle ${\cal L}$ does not have any nontrivial holonomy (i.e.\ delta-function curvature) localized on $P$. This suggests that if the conformal manifold has only weak-coupling singularities, then it might be possible to define a compactification of the manifold ${\cal M}$ and an extension of the bundle ${\cal L}$ on this compactification so that the aforementioned quantization condition continues to hold.

The line bundle ${\cal L}$ and the quantization condition discussed above is reminiscent of the properties of line bundles on the moduli space of supergravity theories \cite{Witten:1982hu}\footnote{When the supergravity theory arises from a compactification of string theory, the supergravity line bundle is related to the aforementioned line bundle on the conformal manifold of the internal 2d ${\cal N}=(2,2)$ worldsheet CFT \cite{Strominger:1990pd, Bershadsky:1993cx}.}. The similarity becomes more precise if we consider the analogue of the Bagger-Witten line bundle restricted on the {\it vacuum manifold} of AdS compactifications of supergravity, which is holographically dual to the conformal manifold of the boundary CFT. We do not explore this direction in this paper, but it would be interesting to investigate it further. It may also be interesting to further explore the structure of the line bundle ${\cal L}$ in the anomaly functional of SCFTs in the presence of a background metric and spacetime-dependent sources for the exactly marginal operators \cite{Gomis:2015yaa}.

Finally, it would be interesting to explore the potential implications and refinements of the K\"ahler-Hodge structure to more general cases of superconformal manifolds, endowed with other types of singularities (e.g.\ orbifold singularities), the role of duality symmetries, and the behavior of the line bundle ${\cal L}$ in these cases (see \cite{Gu:2016mxp} for some related discussions). It would also be interesting to explore potential connections of the K\"ahler-Hodge structure to the CFT distance conjecture \cite{Perlmutter:2020buo}. 

The plan of the paper is as follows: in section \ref{secbackground} we review the necessary background material about superconformal field theories and their marginal deformations. In section \ref{seckahler} we provide a general argument that 4d ${\cal N}=1$ and 3d ${\cal N}=2$ conformal manifolds are K\"ahler. In section \ref{sec4d} we show that ${\cal L}$ is a holomorphic line bundle and compute its curvature deriving the results \eqref{mainresultofpaper}. We collect useful technical details in four appendices at the end of the paper.

\section{Background and setup}
\label{secbackground}

The case of 4d ${\cal N}=2$ and ${\cal N}=4$ theories was discussed in \cite{Papadodimas:2009eu}, where the corresponding special case of the result \eqref{mainresultofpaper} was derived. In the rest of the paper we focus on the remaining theories in $d\geq 3$ with at least 4 supercharges, which can have conformal manifolds \cite{Cordova:2016xhm}, namely 4d ${\cal N}=1$ and 3d ${\cal N}=2$. In both cases, the R-symmetry group of the theory is $U(1)_R$. Our discussion follows the conventions of Refs.\ \cite{Dolan:2002zh,Dolan:2008vc}.

\subsection{Marginal deformations in superconformal field theories}

\subsubsection{4d ${\cal N}=1$}

The symmetry algebra of a 4d ${\cal N}=1$ SCFT possesses two left-chiral Poincar\'e supercharges $Q_a$, $a=1,2$ and two right-chiral Poincar\'e supercharges $\overline{Q}_{\dot{a}}$, where $a, \dot{a}=1,2$ are Weyl spinor indices. It also possesses the corresponding superconformal partners $S^a, \overline{S}^{\dot{a}}$. Under the $U(1)_R$ symmetry group, the supercharges $Q_a$ have R-charge $-1$ and $\overline{Q}_{\dot{a}}$ have $+1$. The R-charge of the superconformal partners is $+1$ for $S^a$ and $-1$ for $\overline{S}^{\dot{a}}$.

We will denote the $\frac{1}{2}$-BPS chiral superconformal primary operators as $\phi_i$ (where $i$ is an index that enumerates them). These operators are annihilated by the supercharges $\overline{Q}_{\dot a}, S^a, \overline{S}^{\dot a}$ and their scaling dimensions $\Delta_i$ and $U(1)_R$ charges $R_i$ obey the shortening conditions $\Delta_i =\frac{3R_i}{2}$. Similarly, the antichiral superconformal primaries $\overline{\phi}_i$ are annihilated by the supercharges $Q_{a}, S^a, \overline{S}^{\dot a}$ and obey the shortening conditions $\Delta_i =-\frac{3R_i}{2}.$

When 4d $\NN=1$ SCFTs have exactly marginal deformations, the corresponding exactly marginal operators are necessarily supersymmetric descendants of chiral (antichiral) primaries $\phi_i$ , ($\overline{\phi}_j$) with scaling dimension $\Delta=3$ and $U(1)_R$-charges $R=2$, ($R=-2$) \cite{Green:2010da}. The chiral and antichiral descendants yield holomorphic and antiholomorphic deformations, respectively, of the form\footnote{The terminology will be justified once we show that the conformal manifolds are complex. In our conventions $\epsilon^{12} = \epsilon^{\dot{1}\dot{2}}=+1$. The minus sign in the definition \eqref{holdef} of $\overline{\cal O}_i$ is necessary to ensure ${\cal O}_i^\dagger = \overline{\cal O}_i$. The overall normalization factor $\frac{1}{4}$ is a convenient convention. The final results are independent of this choice.}
\be
\label{holdef}
{\cal O}_i \equiv \frac{1}{4}\epsilon^{ab} \{Q_a, [Q_b, \phi_i]\}~,~~
\overline{\cal O}_i \equiv - \frac{1}{4} \epsilon^{\dot{a}\dot{b}} \{\overline{Q}_{\dot{a}},[\overline{Q}_{\dot{b}},\overline{\phi}_i]\}
~.
\ee
Exactly marginal deformations of the action take the form\footnote{The overall prefactor of ${1\over (2\pi)^2}$ is a matter of conventions. Changing this prefactor rescales the Zamolodchikov metric by an overall constant, which affects the final formula.
}
\be
S \rightarrow S + {1\over (2\pi)^2 }\delta \lambda^i \int {\cal O}_i + h.c.
\ee
More generally, even if the CFT does not have a Lagrangian, we can think of the change of correlators in conformal perturbation theory as
\be
\nabla_i \langle {\cal O}_1(x_1)...{\cal O}_n(x_n)\rangle ={1\over (2\pi)^2} \int_{\rm ren} d^4z \langle {\cal O}_1(x_1)...{\cal O}_n(x_n){\cal O}_i(z)\rangle
~.
\ee
The covariant derivative on the LHS of this equation involves a connection on the bundle of operators over the conformal manifold and the integral on the RHS is renormalized without allowing operators to collide. Both of these points are reviewed briefly in subsection \ref{connection}.

\subsubsection{3d ${\cal N}=2$}

For spinors in three dimensions we do not need to use dotted and undotted indices. The supercharges are divided into $Q_a$ and $\overline{Q}_a$, with $U(1)_R$ symmetry charges $-1$ and $+1$ respectively. Again, the R-charge of the superconformal partners is $+1$ for $S^a$ and $-1$ for $\overline{S}^{a}$.

Similar to the four-dimensional case of the previous subsection, 3d $\NN=2$ SCFTs can also possess scalar chiral superconformal primary operators annihilated by the super(conformal) charges $\overline{Q}_a, S^a, \overline{S}^a$. Such operators are $\frac{1}{2}$-BPS and their scaling dimensions $\Delta_i$ and $U(1)_R$ charges obey the shortening condition $\Delta_i = R_i$. Antichiral primaries $\overline{\phi}_i$ are annihilated by $Q_a, S^a, \overline{S}^a$ and have scaling dimensions $\Delta_i = -R_i$.

The exactly marginal operators are necessarily supersymmetric descendants of chiral (antichiral) primaries $\phi_i$ ($\overline{\phi}_j$) with $\Delta=2$ and $R=2$ ($R=-2$), of the form
\be
\label{holdef3}
{\cal O}_i \equiv \frac{1}{4} \epsilon^{ab} \{Q_a, [Q_b, \phi_i]\}~, ~~
\overline{\cal O}_i \equiv - \frac{1}{4} \epsilon^{ab} \{\overline{Q}_{a},[\overline{Q}_{b},\overline{\phi}_i]\}
~.
\ee
They yield, respectively, holomorphic and antiholomorphic deformations with corresponding correlation function deformations
\be
\nabla_i  \langle {\cal O}_1(x_1)...{\cal O}_n(x_n)\rangle ={1\over (2\pi)} \int_{\rm ren} d^3z \langle {\cal O}_1(x_1)...{\cal O}_n(x_n){\cal O}_i(z)\rangle~.
\ee

\subsection{Useful superconformal Ward identities}

Here we summarize certain superconformal Ward identities for correlators of marginal operators and correlators of marginal operators with chiral primaries. Both types of identities will be useful in subsequent sections. Their detailed derivation can be found in appendices \ref{appendixward} and \ref{4pointwardproof}.

We begin with a set of straightforward identities that are common in 4d ${\cal N}=1$ and 3d ${\cal N}=2$ SCFTs. They refer to correlation functions involving only holomorphic (antiholomorphic) marginal operators, which vanish. For example,
\be
\label{ward2p}
\langle {\cal O}_i(x) {\cal O}_j(y)\rangle = 0~, 
\ee
\be
\label{ward4p}
\langle {\cal O}_i(x) {\cal O}_j(y){\cal O}_k(z) {\cal O}_l(w)\rangle = 0
\ee
and similarly for their conjugates. 
These identities follow from the fact that ${\cal O} = [Q,...]$ and $[Q,{\cal O}]=0$. Another set of vanishing correlation functions are
\be
\label{mmcp}
\langle {\cal O}_i(x) {\cal O}_j(y) \phi_k(z) \overline{\phi}_l(w) \rangle=0~,
\ee
\be
\label{ward4pm}
\langle {\cal O}_i(x) {\cal O}_j(y){\cal O}_k(z) \overline{\cal O}_l(w)\rangle = 0~.
\ee
These follow by using linear combinations of Ward identities for $Q$'s and $S$'s (see App.\ \ref{appendixward}).

Sometimes we also need to compute correlation functions of the form
\be
\label{identityaa}
\langle {\cal O}_1(\infty) {\cal O}_2(x_2)...{\cal O}_n(x_n)\rangle \equiv \lim_{|x_1|\rightarrow\infty} |x_1|^{2\Delta_1} \langle {\cal O}_1(x_1) {\cal O}_2(x_2)...{\cal O}_n(x_n)\rangle 
~,
\ee
where one of the insertions has been sent to infinity. For example, we will need the Ward identity
\be
\label{ward4der}
\langle \overline{\phi}_i(\infty) \phi_j(y) {\cal O}_k(z)\,\overline{\cal O}_l(w)\rangle = \Box_z \langle \overline{\phi}_i(\infty) {\phi}_j(y) \phi_k(z)\overline{\phi}_l(w)\rangle
~,
\ee
which is valid in this form both in 3d $\NN=2$ and 4d $\NN=1$ SCFTs. Similarly, we will need the correlator of 2 holomorphic and 2 antiholomorphic marginal operators in terms of a corresponding 4-point function of chiral/antichiral primaries. In 4d ${\cal N}=1$ SCFTs we find\footnote{We use the notation $\partial_x^\mu$ or $\partial_\mu^x$ for the partial derivatives with respect to the spacetime variable $x$.}
\be
\label{mainward4}
\begin{gathered}
\langle \overline{\cal O}_i(\infty) {\cal O}_j(y) {\cal O}_k(z) \overline{\cal O}_l(w)\rangle = \Big[(y-z)^2 \Box_y \Box_z + 8 (y-z)^\mu(\partial^y_\mu\Box_z-\partial^z_\mu \Box_y)\\
-32 (\partial_y^\mu \cdot\partial_\mu^z) + 24(\Box_z + \Box_y)\Big]\langle\overline{\phi_i}(\infty)\phi_j(y)\phi_k(z)\overline{\phi}_l(w)\rangle
\end{gathered}
\ee
and in 3d ${\cal N}=2$ SCFTs 
\be
\label{mainward3d}
\begin{gathered}
\langle \overline{\cal O}_i(\infty) {\cal O}_j(y){\cal O}_k(z) \overline{\cal O}_l(w)\rangle = \Big[(y-z)^2\Box_y \Box_z + 6(y-z)^\mu \cdot(\partial^y_\mu \Box_z -\partial^z_\mu \Box_y)\\
-18(\partial^\mu_y\cdot\partial^z_\mu)+12(\Box_y+\Box_z)\Big]\langle \overline{\phi}_i(\infty) \phi_j(y) \phi_k(z) \overline{\phi}_l(w)\rangle
~.
\end{gathered}
\ee
In these expressions, the $\phi/\overline{\phi}$'s are chiral/antichiral primaries of scaling dimension $\Delta=3$ in 4d and $\Delta=2$ in 3d.

\subsection{Connection on the space of operators}
\label{connection}

In conformal perturbation theory, the correlation functions of a CFT deformed by an exactly marginal operator $\OO$ can be expressed in terms of integrated correlation functions of the undeformed CFT. Schematically, the infinitesimal deformation takes the form
\be
\label{confpt}
\delta_{\cal O} \langle \phi_1(x_1)...\phi_n(x_n)\rangle = {1\over (2\pi)^2}\int d^4z \langle{\cal O}(z) \phi_1(x_1)...\phi_n(x_n)\rangle 
~.
\ee
The integral requires regularization at coincident points. In regularizations with a hard space cutoff that excises the coincident points the operators never collide and contact terms do not contribute. Due to this regularization, second order variations by two different marginal operators do not generally commute. It is then natural to think of the LHS of \eqref{confpt} as a {\it covariant} derivative with respect to some connection. The interpretation of this connection is the following: operators with the same quantum numbers mix under conformal perturbation theory. They are to be thought of as sections of vector bundles over the conformal manifold. Conformal perturbation theory naturally defines a connection on this bundle which encodes operator mixing. Alternatively, one can use the operator-state correspondence to reformulate the connection on a conformal manifold as the Berry connection for states of the CFT in radial quantization under adiabatic changes of the exactly marginal couplings \cite{Baggio:2017aww}.

The connection on the space of operators has been discussed in detail for 2d CFTs, in terms of contact terms in \cite{Kutasov:1988xb}, and in terms of regularized integrated correlators in \cite{Ranganathan:1993vj}. A similarly systematic discussion of conformal perturbation theory for higher dimenional CFTs does not exist in the literature. 

For concreteness, consider a set of scalar operators $\phi_K$ with the same quantum numbers. We assume that their common scaling dimension $\Delta$ does not change along the conformal manifold, as is the case for the operators we consider in this paper. These operators can be thought of as sections of a vector bundle ${\cal V}$ over the conformal manifold. On this bundle, one can define an inner product using the 2-point function coefficients
\be
\label{metricbundle}
\langle \phi_K(x) \phi_L(y)\rangle = {g_{KL}(\lambda)\over |x-y|^{2\Delta}}
~.
\ee
The curvature $F$ of the bundle ${\cal V}$ can be computed by a suitably regularized 4-point function of two exactly marginal operators and the operators whose curvature we want to compute
\be
\label{curvdefinition}
(F_{ij})_K^L = {1\over (2\pi)^4}{\rm ren}.\left[\int_{|x|\leq1} d^4 x \int_{|y|\leq 1} d^4y \,\,g^{LM}\,\Big( \langle \phi_M(\infty)\,\, {\cal O}_i(x) {\cal O}_j(y)  \phi_K(0)\rangle   - (x\leftrightarrow y) \Big)\right]
~,\ee
where $g^{LM}$ is the inverse of the metric \eqref{metricbundle}. 

The renormalization scheme employed here is a higher dimensional generalization of the one presented in detail in \cite{Ranganathan:1993vj} (called connection $\overline{c}$ in that paper). First, we consider the $y$ integral in \eqref{curvdefinition}, with $x$ fixed. These is no singularity as $y\rightarrow x$: in the ${\cal O}_i, {\cal O}_j$ OPE the even spin terms cancel because of the antisymmetrization in $x,y$, while the odd spin terms cancel because of the angular integral of $y$ around $x$. Then we consider the potential singularities as $y\rightarrow 0$. Since we are working with external operators $\phi$ whose conformal dimension does not change, there are no logarithmic singularities. Hence, if we define the regularized integral
\be
I(x,\epsilon)\equiv  {1\over (2\pi)^4} \int_{\epsilon\leq |y|\leq 1} d^4y \,\,g^{LM}\,\Big( \langle \phi_M(\infty)\,\, {\cal O}_i(x) {\cal O}_j(y)  \phi_K(0)\rangle   - (x\leftrightarrow y) \Big)
\ee
we expect
\be
I(x,\epsilon) = \sum_{r_i>0} {\alpha_i \over \epsilon^{r_i}} + \widetilde{I}(x) + O(\epsilon)
\ee
where we take $\widetilde{I}$ as the {\it definition} of the finite part of the integral over $y$. 

We then proceed with the integral over $x$ by defining
\be
J(\epsilon) \equiv \int_{\epsilon\leq |x| \leq 1} d^4x \widetilde{I}(x)
\ee
and again expand
\be
J(\epsilon) = \sum_{s^i>0} {\beta_i \over \epsilon^{s_i}} + \widetilde{J}+ O(\epsilon)
\ee
We finally {\it define} the regularized integral in \eqref{curvdefinition} that gives the curvature, by the expression
\be
\label{curvregul}
(F_{ij})_K^L = \widetilde{J}
~.
\ee
Notice that in this scheme no operators ever collide\footnote{We also define the contribution from the region $x=y$ by first cutting-off a ball of size $\epsilon$ and then taking the limit of the integral. As mentioned in the main text, in this case the limit is always finite.}, hence we do not have any possible contributions from contact terms.

For general CFTs, and at a generic point on the conformal manifold, 4-point functions like those appearing in \eqref{curvdefinition} are typically beyond analytic control. In superconformal field theories, when we consider the curvature of operators belonging to short mulitplets, we can often use superconformal Ward identities to simplify these 4-point functions and derive exact non-perturbative properties for the connection.

\subsection{The curvature of the Zamolodchikov metric}

Applying the formula \eqref{curvdefinition} to the bundle of exactly marginal operators, which can be thought of as vectors in the tangent bundle ${\cal TM}$ of the conformal manfiold, we obtain an expression for the Riemann tensor of the Zamolodchikov metric in terms of a doubly integrated 4-point function
\be
\label{riemannf}
R_{ijmn} = {1\over (2\pi)^4} \int_{|x|\leq 1} d^dx \int_{|y| \leq 1} d^d y \, \bigg[ \langle {\cal O}_n(\infty) {\cal O}_i(x) {\cal O}_j(y) {\cal O}_m(0)\rangle - (x\leftrightarrow y)\bigg]
~.
\ee
The expected symmetries of the Riemann tensor follow by using the symmetries of the 4-point functions under global conformal permutations which permute the four points $0,x,y,\infty$ and checking carefully that the regularization scheme does not spoil the symmetries. 

Notice that the formula \eqref{riemannf} gives the Riemann tensor at one point of the conformal manifold in terms of marginal operators ${\cal O}_i$ defined at the same point. Thus, it does not depend on how we select the basis of marginal operators ${\cal O}_i(\lambda)$ in a neighborhood of that point. An arbitrary choice of basis that depends continuously on the exactly marginal couplings $\lambda$ can be made.
 
Now, let us assume that we chose a specific basis of marginal operators  ${\cal O}_i(\lambda)$ in a neighborhood of a point. We compute the 2-point functions
\be
\label{anothereq}
\langle {\cal O}_i(x) {\cal O}_j(y) \rangle = {G_{ij}(\lambda) \over |x-y|^8}
~.
\ee
In general, we {\it can not} directly interpret the matrix $G_{ij}(\lambda)$ as a (Zamolodchikov) metric on the conformal manifold in a coordinate frame. In particular, if we try to compute the Riemann tensor from \eqref{anothereq} by using the usual differential geometry formulae, which give the Christoffel symbols and the Riemann tensor by combinations of derivatives of the metric, it {\it will not} agree with the computation from the 4-point function \eqref{riemannf}. For example, in any family of CFTs we can select a basis of marginal operators in a neighborhood of a point such that $G_{ij}(\lambda) = \delta_{ij}$ everywhere in that neighborhood, which would erroneously suggest that the Riemann tensor is zero. 

The geometric interpretation of the above observations is clear. In general, a choice of a basis of marginal operators ${\cal O}_i(\lambda)$ should be thought of as a choice of a basis section on the tangent bundle. This choice will not be related always to a coordinate system. The sections ${\cal O}_i(\lambda)$ of marginal operators is a set of linearly independent vector fields on the conformal manifold, which are not in general {\it integrable}, i.e.\ they can not be thought of as dual to coordinates on the manifold.

This issue does not arise when we define marginal operators in terms of the variation of certain parameters in the Lagrangian, since then they are automatically integrable. However, from an abstract CFT point of view, that does not employ the use of a Lagrangian, we need to check independently whether we are working with integrable marginal operators. One necessary, though not sufficient, condition for a choice of marginal operators ${\cal O}_i(\lambda)$ to correspond to tangent vectors in the system of coordinates $\lambda^i$ is that the Riemann tensor as computed by thinking of \eqref{anothereq} as the actual metric agrees with that from \eqref{riemannf}.

\subsection{Vector bundles of chiral primaries}

Since $U(1)_R$ charge conservation prevents operator mixing between chiral primaries of different charge, for each value of the R-charge we have a corresponding vector sub-bundle ${\cal V}_R$. 

As we will discuss later, the conformal manifolds of the theories we are considering are complex, K\"ahler and the bundles ${\cal V}_R$ are holomorphic vector bundles. In 4d SCFTs with extended supersymmetry ($\NN=2, 4$) the bundles of half-BPS operators exhibit additional structure. In $\NN=4$ theories the corresponding bundles are flat (a property which is related to a non-renormalization theorem for 3-point functions \cite{Lee:1998bxa,DHoker:1998vkc,DHoker:1999jke,Intriligator:1998ig,Intriligator:1999ff,Eden:1999gh,Petkou:1999fv,Howe:1999hz,Heslop:2001gp,Baggio:2012rr}). In $\NN=2$ theories there are two types of half-BPS operators: Higgs-branch and Coulomb-branch, \cite{Dolan:2002zh}. The bundles of the former type are also flat \cite{Niarchos:2018mvl}, while the bundles of the latter type exhibit non-trivial curvature, which obeys the $tt^*$ equations \cite{Papadodimas:2009eu,Baggio:2014ioa}. SCFTs with four Poincar\'e supercharges (in 3d or 4d) are expected to have bundles of half-BPS chiral primary operators with non-trivial curvature as well, but it is not known if these bundles exhibit a structure akin to the $tt^*$ geometry. The $tt^*$ geometry of half-BPS operators in 2d $\NN=(2,2)$ theories was originally discussed in \cite{Cecotti:1991me}.    

Of particular interest in this paper will be the bundle of chiral primaries in 4d $\NN=1$ and 3d $\NN=2$ SCFTs, whose descendants are marginal operators (i.e.\ chiral primaries of $\Delta=3$ in 4d and $\Delta=2$ in 3d). We denote these bundles as ${\cal V}$ (and $\overline{\cal V}$ for the corresponding anti-chiral primaries). 

\subsection{Supercharge line bundle ${\cal L}$}

We need one more ingredient: the line bundle ${\cal L}$ encoding the operator mixing between the supercharges. We start by considering the supercurrents. In 4d ${\cal N}=1$ SCFTs, the supercurrents are conformal primary operators $G_{\mu a}, \overline{G}_{\mu\dot{a}}$ with scaling dimension $\Delta={7\over 2}$ and spin $(1,{1\over 2})$ and $({1\over 2},1)$ respectively. Since they are the only operators with these quantum numbers (otherwise SUSY would be enhanced) the only possible mixing that we can have is a phase rotation of the form
\be 
\label{scphase}
G_{\mu a} \rightarrow e^{i \theta }G_{\mu a} \qquad,\qquad \overline{G}_{\mu\dot{a}} \rightarrow e^{-i \theta} \overline{G}_{\mu\dot{a}}
~.
\ee 
As we will see, conformal perturbation theory implies that there is indeed a nontrivial phase rotation under conformal perturbation theory on the conformal manifold. This means that the supercurrents $G$ should be associated to a line bundle ${\cal L}$ over the conformal manifold (and the conjugate $\overline{G}$ to a line bundle $\overline{\cal L}$).

Recall that the supercharges (and superconformal charges) are integrals of the supercurrents
\be
Q_a = \int d^3x \,G_{0,a} \qquad \overline{Q}_{\dot{a}} = \int d^3x \,\overline{G}_{0,\dot{a}}
\ee
and
\be
S_{a} = \int d^3x \,x^{a \dot{a}} \,\overline{G}_{0,\dot{a}} \qquad \overline{S}_{\dot{a}} = \int d^3x \,x^{a \dot{a}} \,G_{0,a}
\ee  
so the phase rotation \eqref{scphase} implies the following rotation of the supercharges and their superconformal partners
$$
Q_{a} \rightarrow e^{i \theta} \, Q_a~~ ,~~ \overline{Q}_{a} \rightarrow e^{-i \theta} \,\overline{Q}_a~,
$$
$$
S_{a} \rightarrow e^{-i \theta} \,S_a~~ ,~~ \overline{S}_{\dot{a}} \rightarrow e^{i \theta} \,\overline{S}_{\dot{a}}
~,
$$
which corresponds to an automorphism of the ${\cal N}=1$ superconformal algebra\footnote{For 4d SCFTs with extended supersymmetry we have $
Q_{a}^I \rightarrow e^{i \theta} \, Q_a^{I}~,~ \overline{Q}_{a}^I \rightarrow e^{-i \theta} \,\overline{Q}_a^{I}~,
$
$
S_{a}^I \rightarrow e^{-i \theta} \,S_a^{I}~,~ \overline{S}_{\dot{a}}^I \rightarrow e^{i \theta} \,\overline{S}_{\dot{a}}^{I}~.
$
For ${\cal N}=1,2$ this automorphism is inner (it can be generated by R-charge rotation) while for ${\cal N}=4$ it is outer.}. Completely analogous statements hold for the 3d ${\cal N}=2$ SCFTs.

The conclusion is that the supercharges and superconformal partners should be thought of as associated to a line bundle ${\cal L}$ over the conformal manifold, whose curvature is determined by the dynamics of the CFT. Similar results hold for 2d SCFTs, see for example \cite{Distler:1992gi} for a review in ${\cal N}=(2,2)$ 2d theories. For completeness, we notice that for 2d ${\cal N}=(4,4)$ theories the corresponding bundles of supercurents have $SU(2)_L\otimes SU(2)_R$ structure \cite{deBoer:2008ss}.

\section{K\"ahler structure}
\label{seckahler}

In this section we show that the superconformal manifolds of interest are complex and K\"ahler. An argument for the K\"ahler property of 4d ${\cal N}=1$ superconformal manifolds was given in \cite{Asnin:2009xx} based on the possible forms of contact terms that can be written in superspace. Here we recast the argument in a language which is more suitable to a scheme of conformal perturbation theory that does not involve directly contact terms. In addition, we generalize the proof in \cite{Asnin:2009xx} in two ways. First, in contrast to \cite{Asnin:2009xx} we do not {\it assume} the existence of local complex coordinates but instead we prove it. Second, our proof also applies to 3d ${\cal N}=2$ theories. Finally, we provide a check of K\"ahlerity by explicitly computing the Riemann tensor at a generic point of the conformal manifold and show that it has the expected properties of a K\"ahler manifold.

\subsection{Proof of K\"ahlerity}
\label{kaehlerproof}

At each point on the conformal manifold supersymmetry divides the marginal operators into descendants of chiral primaries and antichiral primaries \eqref{holdef}, \eqref{holdef3}, hence the tangent space is split pointwise into holomorphic and antiholomorphic subspaces. This defines an almost complex structure on the conformal manifold. The integrability of this complex structure, and relatedly the existence of local complex coordinates, is not obvious from this argument and will be proven below.

The Zamolodchikov metric is Hermitian with respect to this almost complex structure: the only non-vanishing components are of the form $G_{i\overline{j}}$, since the Ward identities \eqref{ward2p} imply that 
\beq
G_{ij} = G_{\overline{i} \,\overline{j}} = 0
~.
\eeq 
The arguments so far demonstrate that the conformal manifold is an {\it almost hermitian manifold}.

To show that the almost complex structure is integrable and that the metric is K\"ahler we proceed as follows. Consider the following vector bundles over the conformal manifold which were all introduced in section 2: the tangent bundle ${\cal TM}$, the supercharge line bundle ${\cal L}$ and the vector bundles ${\cal V},\overline{\cal V}$ of chiral, antichiral primaries whose descendants are the exactly marginal operators.

Each of these bundles is equipped with a connection, which is determined by conformal perturbation theory. Suppose that the conformal manifold has real dimension $2n$. The tangent bundle ${\cal TM}$ has (local) holonomy which is generically $SO(2n)$. The vector bundles ${\cal V},\overline{\cal V}$ have holonomy which is generically $U(n)$. The line bundle ${\cal L}$ has holonomy $U(1)$.

In addition, the holonomy of the tangent bundle ${\cal TM}$ splits as
\be
{\cal T M}  = ({\cal L}^2 \otimes {\cal V})\oplus (\overline{\cal L}^{2}\otimes \overline{\cal V})~,
\ee
or
\be
{\cal T}{\cal M} = {\cal T}_{\rm hol}{\cal M}\oplus {\cal T}_{\rm anti-hol}{\cal M}~,
\ee
where we defined the holomorphic/antiholomorphic split as
$
{\cal T}_{\rm hol}{\cal M} = {\cal L}^2\otimes {\cal V}~,
$
$
{\cal T}_{\rm anti-hol}{\cal M} = \overline{\cal L}^{2}\otimes \overline{\cal V}
~.
$

This implies that the holonomy of the tangent bundle ${\cal TM}$ is reduced to $U(n)$. Then, Berger's classification theorem \cite{BSMF_1955__83__279_0} implies that the conformal manifold ${\cal M}$ is K\"ahler. In particular, it guarantees the existence of local complex coordinates compatible with the almost hermitian structure.

An important point in this argument is that chiral primaries ${\cal V}$ and antichiral primaries $\overline{\cal V}$ can not mix due to charge conservation. Another important point is that when considering the exactly marginal operators ${\cal O}_i = \frac{1}{4} \epsilon^{ab} \{Q_a,[Q_b,\phi_i]\}$, the connection for ${\cal O}_i$ is the sum of the connection for $\phi_i$ plus two times that for the $Q$'s. We confirm that this is indeed the case in appendix \ref{appadd}.

\subsection{A check: direct computation of the curvature}

In this subsection we provide a check of the previous formal argument that the manifold is K\"ahler by confirming that the Riemann tensor in 4d ${\cal N}=1$ SCFTs satisfies at generic points the expected properties of K\"ahlerity. Exactly analogous results can be derived for 3d ${\cal N}=2$ theories.

For a K\"ahler manifold we expect the following components of the Riemann tensor to be zero
$$
R_{ijkl} \quad R_{ijk\overline{l}} \quad R_{ij\overline{l}k}\quad R_{ij\overline{k}\,\overline{l}}
~,
$$
and their complex conjugate, as well as
$$
R_{i\overline{j}kl} \quad R_{i\overline{j}\overline{k}\,\overline{l}} \quad R_{\overline{j}ikl} \quad R_{\overline{j}i\overline{k}\, \overline{l}}
~.
$$

Taking into account the symmetry properties of $R$, as well as the behavior under complex conjugation we have the following independent vanishing conditions that need to be verified
\be
\label{riemann1}
R_{ijkl} =0
~,
\ee
\be
\label{riemann2}
R_{ijk\overline{l}}=0
~,
\ee
\be
\label{riemann3}
R_{ij\overline{k}\,\overline{l}}=0
~.
\ee
For a K\"ahler manifold, the Riemann tensor also obeys an additional symmetry property
\be
R_{i\overline{j}k\overline{l}} = R_{i\overline{l}k\overline{j}}
~,
\ee
which is not true for general Riemannian manifolds. However, this symmetry follows from the 1st Bianchi identity of the Riemann tensor combined with the previous vanishing statements.

The vanishing of the components \eqref{riemann1} and \eqref{riemann2} follows from the vanishing of the 4-point functions \eqref{ward4p} and \eqref{ward4pm} by Ward identities. It is a little harder to prove the vanishing of \eqref{riemann3}. The relevant 4-point function is
\be
\label{kahlercor}
\langle {\cal O}_i(\infty) \overline{\cal O}_k(x) \overline{\cal O}_l(y) {\cal O}_j(0)\rangle
\ee
This is {\it not} identically zero, but a vanishing curvature arises after integration in the relevant formula
$$
R_{i j,\overline{k}\,\overline{l}} = {1\over (2\pi)^4} \int_{|x|\leq 1} d^4x \int_{|y|\leq 1} d^4 y \left[\langle {\cal O}_i(\infty) \overline{\cal O}_k(x) \overline{\cal O}_l(y) {\cal O}_j(0)\rangle- (x\leftrightarrow y)\right]
~.
$$
This double integral can be drastically simplified by integration by parts using the Ward identity \eqref{mainward4}. After a straightforward analysis, that we present in appendix \ref{kahlercurv}, we find that the integral vanishes and thus \eqref{riemann3} is satisfied.

\section{K\"ahler-Hodge structure}
\label{sec4d}

In this section we compute the curvature of the (supercharge) line bundle ${\cal L}$ and show that the conformal manifold is not only K\"ahler, but in fact K\"ahler-Hodge. We start by first showing that ${\cal L}$, ${\cal V}$ are holomorphic vector bundles.

\subsection{Proof that ${\cal L}, {\cal V}$ are holomorphic line bundles}

Consider the purely holomorphic components of the curvature of ${\cal L}$. Since
$$
{\cal T}_{\rm hol}{\cal M}= {\cal L}^2 \otimes {\cal V}
$$
we have
\be
\label{curvadd}
R^l_{ijk}=2 \delta^l_k F^{\cal L}_{ij} +(F^{{\cal V}}_{ij})_k^l 
~.
\ee
Hence, the curvature of the bundle ${\cal L}$ can be computed via the difference of the curvature of ${\cal T}_{\rm hol}{\cal M}$ and ${\cal V}$. 
For the computation of the holomorphic curvature components $(F^{{\cal V}}_{ij})_k^l$ of the bundle of chiral primaries ${\cal V}$ we need to use in \eqref{curvdefinition} the 4-point function
\be
\langle \overline{\phi}_m(z) {\cal O}_i(x) {\cal O}_j(y) \phi_k(w)\rangle
~,
\ee
which vanishes identically due to the Ward identities \eqref{mmcp}. Consequently,
 \be
 (F^{{\cal V}}_{ij})_k^l=0
 \ee
 Similarly, we can show that
 \be
 (F^{{\cal V}}_{\overline{i}\,\overline{j}})_k^l=0
 ~.
 \ee
 Altogether, we conclude that ${\cal V}$ is a holomorphic line bundle, see also \cite{Papadodimas:2009eu}.
 
In the previous section we argued that  $R^l_{ijk}$ vanishes due to the Ward identity \eqref{ward4pm}. Hence, from  
\eqref{curvadd} we find that 
\be
\label{holbundle1}
F_{ij}^{\cal L} = 0
~.
\ee
Similarly,
\be
\label{holbundle2}
F_{\overline{i}\,\overline{j}}^{\cal L} = 0
~.
\ee
As a result, we have established that ${\cal L}$ is a holomorphic line 
bundle over the conformal manifold. 

This result can also be obtained by directly computing the purely holomorphic components of the curvature of the supercurrent operator using Eq.\ \eqref{curvdefinition}. In that case we need to consider the 4-point function
\be 
\label{directg}
\langle  \overline{ G}_{\nu,\dot{a}}(\infty) {\cal O}_i(x) {\cal O}_j(y)\, G_{\mu,a}(0)\rangle
~.
\ee 
Using the fact that $G_{\mu,a} = [Q_a,J_\mu]$, where $J_\mu$ is the R-current, and the appropriate superconformal Ward identities it is easy to show that \eqref{directg} vanishes. This is an alternative proof of \eqref{holbundle1} and similarly \eqref{holbundle2}.

\subsection{Mixed components of the curvature of ${\cal L}$ in 4d ${\cal N}=1$}

For the mixed components of the curvature
\be
\label{sumcurv}
R^l_{k,i\overline{j}} =  2 \delta_k^l F^{\cal L}_{i\overline{j}}+(F^{\cal V}_{i\overline{j}})_k^l
\ee
To compute the LHS we need to apply the curvature formula \eqref{curvdefinition}  to the 4-point function
$$
G^{\overline{m}l}\langle \overline{\cal O}_m(\infty) {\cal O}_i(x) \overline{\cal O}_j(y) {\cal O}_k(z)\rangle   $$
We start with the Ward identity \eqref{mainward4} 
\be
\label{aa}
\begin{gathered}
\langle \overline{\cal O}_m(\infty) {\cal O}_i(x) \overline{\cal O}_j(y) {\cal O}_k(z)\rangle   =
\Big[ (x-z)^2 \Box_x \Box_z + 8 (x-z)\cdot(\partial_x\Box_z-\partial_z \Box_x)\\
-32 (\partial_x \cdot\partial_z) + 24(\Box_z + \Box_x)\Big] \langle\overline{ \phi}_m(\infty)\phi_i(x)\overline{\phi}_j(y)\phi_k(z)\rangle
\end{gathered}
\ee
and take the limit $z\rightarrow 0$. To apply the usual curvature formula \eqref{curvdefinition} it is convenient to eliminate the $z$-derivatives. This can be done using the ordinary translation Ward identity
\be
\label{dztodxy}
\partial_\mu^z \langle\overline{ \phi}_m(\infty)\phi_i(x)\overline{\phi}_j(y)\phi_k(z)\rangle = -(\partial_\mu^x+\partial_\mu^y)\langle\overline{ \phi}_m(\infty)\phi_i(x)\overline{\phi}_j(y)\phi_k(z)\rangle
~.
\ee
Implementing \eqref{dztodxy} into \eqref{aa}, and sending $z\rightarrow 0$, we obtain 
\be
\label{onemore}
\begin{gathered}
\langle \overline{\cal O}_m(\infty) {\cal O}_i(x) \overline{\cal O}_j(y) {\cal O}_k(0)\rangle   =\\=
\Big\{x^2 \Box_x (\partial_x + \partial_y)^2 + 8 x\cdot(\partial_x(\partial_x + \partial_y)^2+(\partial_x + \partial_y) \Box_x)\\
+32 (\partial_x \cdot(\partial_x + \partial_y)) + 24[(\partial_x + \partial_y)^2 + \Box_x]\Big\}
\langle\overline{ \phi}_m(\infty)\phi_i(x)\overline{\phi}_j(y)\phi_k(0)\rangle
~.
\end{gathered}
\ee

To proceed further we notice that the 2-point functions of chiral primaries with scaling dimension $\Delta=3$ and the 2-point functions of marginal operators 
\be
\langle \phi_i(x) \overline{\phi}_j(y) \rangle = {g_{i\overline{j}} \over |x-y|^6}  \qquad,\qquad \langle {\cal O}_i(x) \overline{\cal O}_j(y) \rangle = {G_{i\overline{j}} \over |x-y|^8}
\ee
are related by Ward identities as
\be
\langle {\cal O}_i(x) \overline{\cal O}_j(y) \rangle = 
\Box_x \langle \phi_i(x) \overline{\phi}_j(y) \rangle
~,
\ee
which implies that $G_{i\overline{j}} = 24 
g_{i\overline{j}}$ and $G^{\overline{m}l} = {1\over 24} 
g^{\overline{m}l}$.

We then multiply \eqref{onemore} by $G^{\overline{m}l}$, we split off the last term and use the Ward identity \eqref{ward4der} to write
\be
\label{onemore2}
\begin{gathered}
G^{\overline{m}l}\langle \overline{\cal O}_m(\infty) {\cal O}_i(x) \overline{\cal O}_j(y) {\cal O}_k(0)\rangle =  W(x,y)+g^{\overline{m}l}
\langle\overline{ \phi}_m(\infty){\cal O}_i(x)\overline{\cal O}_j(y)\phi_k(0)\rangle \end{gathered}
\ee
where
\be
\begin{gathered}
W(x,y) \equiv 
\Big\{x^2 \Box_x (\partial_x + \partial_y)^2 + 8 x\cdot(\partial_x(\partial_x+\partial_y)^2+(\partial_x + \partial_y) \Box_x)\\
+32 (\partial_x \cdot(\partial_x + \partial_y)) + 24(\partial_x + \partial_y)^2\Big\} 
G^{\overline{m}l}\langle\overline{ \phi}_m(\infty)\phi_i(x)\overline{\phi}_j(y)\phi_k(0)\rangle~.
\end{gathered}
\ee
The last term in Eq.\ \eqref{onemore2} precisely reproduces the curvature of the chiral primaries, that is the second term in \eqref{sumcurv}. Hence, we find that the curvature of the line bundle ${\cal L}$ is
\be
\label{curveeqagain}
F^{\cal L}_{i\overline{j}}  \delta^{l}_k={1\over 2} {1\over (2\pi)^4}\int_{|x|\leq 1} d^4x \int_{|y|\leq 1} d^4y\, \Big(W(x,y) -W(y,x) \Big) 
~.
\ee
Our next goal is to try to recast $W$ as a double total-derivative to facilitate the computation of the integral on the RHS of Eq.\ \eqref{curveeqagain}.

First, we rewrite $W$ as
\be
\label{onemore3}
\begin{gathered}
W=\Big\{\Box_x x^2 \Box_x +  \Box_x \Box_y x^2  +2  (\partial_x\cdot \partial_y)  x^2 \Box_x\\
+12 (x\cdot \partial_x) \Box_x+ 4(x \cdot\partial_x) \Box_y +16 (x\cdot \partial_x) (\partial_x \cdot \partial_y)  + 4 (x\cdot \partial_y) \Box_x
\\+48 \Box_x + 80(\partial_x\cdot\partial_y)+16\Box_y \Big\}G^{\overline{m}l}
\langle\overline{ \phi}_m(\infty)\phi_i(x)\overline{\phi}_j(y)\phi_k(0)\rangle~.
\end{gathered}
\ee
Some of the terms above have only $x$-derivatives. In order to be able to do a double partial integration in \eqref{curveeqagain} we want to convert some of them to $y$-derivatives. To do that we can use the following conformal Ward identity, which is true for scalar operators of conformal dimension $\Delta=3$,\footnote{\label{footnoteid}This identity can be proved easily by direct computation after expressing the 4-point function in terms of the conformally-invariant cross-ratios (see Eq.\ \eqref{crossratios} below).}
\be
\Box_x \langle\overline{ \phi}_m(\infty)\phi_i(x)\overline{\phi}_j(y)\phi_k(0)\rangle = {1\over x^2 y^2} \Box_y \Big(y^4\langle\overline{ \phi}_m(\infty)\phi_i(x)\overline{\phi}_j(y)\phi_k(0)\rangle \Big)
\ee
to recast \eqref{onemore3} as
\be
\begin{gathered}
W= 
\big\{\Box_x \Box_y y^2 +4\Box_x \partial_k^y y^k +  \Box_x \Box_y x^2  +2  (\partial_x\cdot \partial_y)  x^2 \Box_x\\
+12 (\partial_x\cdot x) (\Box_y {y^2\over x^2} + 4 \partial_k^y {1\over x^2}y^k)+ 4(\partial_x\cdot x) \Box_y +16 (\partial_x \cdot \partial_y) (\partial_x\cdot x)  + 4 \Box_x(x\cdot \partial_y) \\-8(\partial_x\cdot\partial_y)\big\}
G^{\overline{m}l}\langle\overline{ \phi}_m(\infty)\phi_i(x)\overline{\phi}_j(y)\phi_k(0)\rangle
~.
\end{gathered}
\ee
This has allowed us to write $W$ a double total-derivative with respect to both $x$ and $y$
\be
W(x,y) = \partial_\mu^y \partial_\nu^x H^{\mu\nu}(x,y)
~,
\ee
where
\be
\label{hformula}
\begin{gathered}
 H^{\mu\nu}(x,y) = 
\Big\{\partial^\mu_y \partial^\nu_x y^2+ 4 y^\mu \partial^\nu_x + \partial^\mu_y \partial^\nu_x x^2 + 2 \eta^{\mu\nu} x^2 \Box_x
+ 12 x^\nu \left(\partial_y^\mu{y^2 \over x^2} + 4 {y^\mu \over x^2}\right)\\
+4\partial^\mu_y x^\nu +16 \eta^{\mu\nu} (\partial_x\cdot x) + 4 \partial^\nu_x x^\mu -8 \eta^{\mu\nu}\Big\}
G^{\overline{m}l}\langle\overline{ \phi}_m(\infty)\phi_i(x)\overline{\phi}_j(y)\phi_k(0)\rangle
~.
\end{gathered}
\ee
Moreover,
\be
W(y,x) = \partial_\mu^x \partial_\nu^y H^{\mu\nu}(y,x) = \partial_\mu^y \partial_\nu^x H^{\nu\mu}(y,x)
~.
\ee

Consequently, we need to compute the quantity
\be
\begin{gathered}
F^{\cal L}_{i\overline{j}}  \delta^{l}_k={1\over 2} {1\over (2\pi)^4}\int_{|x|\leq 1} d^4x  \int_{|y|\leq 1} d^4y\, (W(x,y) -W(y,x)) 
\\
={1\over 2} {1\over (2\pi)^4}\int_{|x|\leq 1} d^4x \int_{|y|\leq 1}d^4y\,\,\, \partial_\mu^y\partial_\nu^x( H^{\mu\nu}(x,y)-H^{\nu\mu}(y,x))
~.
\end{gathered}
\ee
Doing the integration by parts we find\footnote{To arrive at this expression we have checked that there are no finite contributions from the limits $x\to 0$ and $y\to 0$, according to the regularization method given in equations \eqref{curvdefinition}-\eqref{curvregul}. Such potential contributions either cancel out automatically or they have to involve scalar operators with scaling dimension 2 or 4 in the OPE of a chiral and an anti-chiral operator. It can be argued that these are absent: $i)$ The argument that scaling dimension 2 operators cannot contribute appears below, see equation \eqref{scalarzeroa}. $ii)$ At generic points of the conformal manifold, scalar scaling-dimension-4 operators would have to be exactly marginal, namely descendants of chiral primaries. Since the corresponding three-point functions vanish, as a consequence of superconformal Ward identities, such operators cannot contribute.}
\be
\label{onemore4a}
F^{\cal L}_{i\overline{j}}  \delta^{l}_k={1\over 2} {1\over (2\pi)^4} \lim_{r\rightarrow 1^-}\int_{|y|=1}d\Omega_y\,\,\int_{|x|=r} d\Omega_x\,\,|x|^2|y|^2 y_\mu x_\nu   ( H^{\mu\nu}(x,y)-H^{\nu\mu}(y,x))
~.
\ee
 Furthermore, in the limit $r\rightarrow 1^-$, the contribution to the integral from angular regions on the spheres $|x|=1,|y|=1$, which have finite distance, cancels after antisymmetrization. The only possible contributions come from the region $x\approx y$ on the spheres. We need to evaluate this contribution {\it before} taking the $r\rightarrow 1^-$ limit. To compute it we can use the OPE between the operators at $x,y$.

Hence, to evaluate these integrals it is convenient to perform a conformal block expansion of the 4-point functions in the channel $(i\overline{j})\rightarrow (k\overline{l})$. We denote by ${\cal O}$ the primary exchanged operator of dimension $\Delta$ and spin $s$\footnote{Since the external operators are scalars, in the intermediate channel we will only have operators with Lorentz spin $j_L=j_R$. We define $s=2j_L$.}. The relevant 4-point functions can be expanded as
\be
\langle \overline{\phi}_m(\infty) \,\phi_i(x)\,   \overline{\phi}_j(y)\,\, \phi_k (0)\rangle={1\over |x-y|^6} \sum_{\Delta,s}{C_{i\overline{j}}^{\cal O} C_{{\cal O} k \overline{m}}}\,\, g_{\Delta,s}(u,v)
~,
\ee
where $g_{\Delta,s}(u,v)$ is the conformal block and $u,v$ the cross-ratios given by
\be
\label{crossratios}
u = {|x-y|^2 \over |x|^2}\qquad v = {|y|^2 \over |x|^2}
~.
\ee
For the second term in \eqref{onemore4a} we have a similar expansion interchanging $x,y$, which implies $u\rightarrow u/v$, $ v\rightarrow 1/v$. Hence, we have
\be
\langle \overline{\phi}_m(\infty) \,\phi_i(y)\,   \overline{\phi}_j(x)\,\, \phi_k (0)\rangle={1\over |x-y|^6} \sum_{\Delta,s}{C_{i\overline{j}}^{\cal O} C_{{\cal O} k \overline{m}}}\,\, g_{\Delta,s}(u/v,1/v)
~.
\ee
Since conformal blocks obey the identity
\be
g_{\Delta,s}(u/v,1/v) = (-1)^s g_{\Delta,s}(u,v)~,
\ee
we can write the final result for the curvature as
\be
F ={1\over 2} G^{\overline{m}l} \times \sum_{\Delta,s} C_{i\overline{j}}^{\cal O} C_{{\cal O} k \overline{m}} \,\, X_{\Delta,s}
~,
\ee
where we have defined the kinematic quantity
\be
\begin{split}
& X_{\Delta,s} \equiv {1\over (2\pi)^4}\lim_{r\rightarrow 1^-} \int_{|x|=r} d\Omega_3^x \int_{|y|=1}d\Omega_3^y\, |y|^2 |x|^2  \times\cr
&
 \times
\bigg\{\big[(x\cdot \partial_x)(y\cdot \partial_y) y^2 +4y^2(x\cdot \partial_x)
+(x\cdot\partial_x)(y\cdot \partial_y) x^2 + 2 (x\cdot y) x^2 \Box_x \cr 
&+12 (y\cdot \partial_y) y^2 +48 y^2+4 x^2 (y\cdot\partial_y)+16 (x\cdot y)(x\cdot\partial_x + 4)
+4 (x\cdot\partial_x)(x\cdot y)-8 (x\cdot y)\big]{g_{\Delta,s}(u,v)\over |x-y|^6} \cr
 &
 -(-1)^s\times\left[\text{same string as above with $x,y$ interchanged acting on}\right]\, { g_{\Delta,s}(u,v)\over |x-y|^6}\bigg\}~.
\end{split}
\ee
After some work in Mathematica we find that the only non-zero $X_{\Delta,s}$ are those corresponding to intermediate exchanged operators ${\cal O}$ with 
\begin{enumerate}
 \item $\Delta=2,s=0$ with coefficient $X_{2,0}=3$.
 \item $\Delta=3,s=1$ with coefficient $X_{3,1}=6$.
\end{enumerate}
So finally we find that the curvature is
\be
\label{final51}
F_{i\overline{j}}\delta_k^l ={1\over 2}G^{\overline{m}l} 
\left (3\sum_n C_{i\overline{j}}^{A_n} C_{A_n k \overline{m}} + 6 \sum_n C_{i\overline{j}}^{J_n} C_{J_n k \overline{m}} \right)
~,
\ee
where the first sum runs over scalar primaries $A_n$ of $\Delta =2$, $s=0$ and the second over conserved currents $J_n$ of $\Delta=3$, $s=1$.
\vskip10pt

\vskip10pt

\noindent {\it Scalar contribution}
 
\vskip10pt

Consider a neutral scalar primary $A$ with scaling dimension $\Delta=2$. This belongs to a short multiplet \cite{Cordova:2016emh} and the second superconformal descendant of the form $J_{a\dot{a}}=[Q_a,[\overline{Q}_{\dot{a}}, A]]$  is a conserved flavor current. At a generic point on the conformal manifold the marginal operators must be neutral under flavor currents \cite{Green:2010da}. This implies that $\phi_i$ must be neutral under $J$, hence
\be 
\langle \phi_i \overline{\phi}_j J_{a\dot{a}} \rangle = 0
~.
\ee
Inserting the expression $J_{a\dot{a}}=[Q_a,[\overline{Q}_{\dot{a}}, A]]$ and using appropriate superconformal Ward identities we deduce that
\be
\label{scalarzeroa}
\langle \phi_i \overline{\phi}_j A\rangle = 0
~.
\ee
As a result, the scalar contribution in \eqref{final51} vanishes.

\vskip10pt

\noindent {\it Spin 1 contribution}

\vskip10pt

First, we consider a flavor (non-R) current $J^F_\mu$. At a generic point on the conformal manifold the marginal operator must have zero charge under $J^F_\mu$. Since the supercharges $Q_a$ are uncharged under $J^F_\mu$, this implies that the chiral primaries $\phi_i$ must have zero charge under $J^F_\mu$ and hence $J_\mu^F$ cannot appear in the $\phi_i \overline{\phi}_j$ OPE. This argument does not apply to the R-current, since $Q_a$ is charged under it.

To compute the contribution of the R-current we proceed as follows. Suppose that the 2-point function of the R-current is normalized as
\be
\langle J_\mu(x) J_\nu(0)\rangle = {C_V \over |x|^6}
\left( \eta_{\mu\nu} - 2\frac{x_\mu x_\nu}{x^2} \right)~.
\ee
We also have the OPEs
\be
J_\mu(x) \phi_i(0) =i  {R_i\over 2\pi^2} {x^\mu \over |x|^4} \phi_i(0)+...~,
\ee
\be
\phi_i(x) \overline{\phi}_j(y) = -i{R_i \over 2\pi^2 C_V} g_{i\overline{j}} {(x-y)^\mu \over |x-y|^4} J_\mu(y)+...~.
\ee
These imply that in the limit $x\rightarrow y$
\be
\label{fourexpand}
\langle\overline{\phi}_m(\infty)\phi_i(x)\overline{\phi}_j(y)\phi_k(0)\rangle =  {R_i R_k\over 4\pi^4 C_V} g_{i\overline{j}} g_{k\overline{m}} {(x-y)\cdot y \over |x-y|^4 |y|^4}+...~.
\ee 
In our case we have $R_i=R_k=2$. Also for 4d ${\cal N}=1$ SCFTs we have  $C_V={4\over \pi^4}  c$. Hence, comparing \eqref{fourexpand} to the conformal block expansion we find

\be
\label{contributioncurrent}
C^J_{i\overline{j}}C_{Jk\overline{m}}  = -{1 \over 4 c} g_{i\overline{j}} g_{k\overline{m}}~.
\ee

\vskip20pt
\noindent {\it Final expression for the curvature}
\vskip20pt

Implementing \eqref{scalarzeroa} and \eqref{contributioncurrent} in \eqref{final51} we finally obtain
\be
 F^{\cal L}_{i \overline{j}} \delta^l_k= {1\over 2}G^{\overline{m}l}\times 6\times \left(-{1\over 4c} g_{i\overline{j}} g_{k\overline{m}}\right)
~,
\ee
that is
\be
F^{\cal L}_{i \overline{j}} = -{1\over 32 c}g_{i\overline{j}}
~.
\ee
In terms of the Zamolodchikov metric $G_{i\overline{j}}= 24 g_{i\overline{j}}$
\be
\label{finalcurvature1}
F^{{\cal L}}_{i\overline{j}} =  -{1\over 768 c} G_{i\overline{j}}
~.
\ee

\subsection*{Comparison with results in ${\cal N}=2$ and ${\cal N}=4$ SCFTs}

A similar computation for the curvature of the line bundle $\LL$ in 4d ${\cal N}=2$ SCFTs was performed in \cite{Papadodimas:2009eu}. There it was found that the corresponding curvature is $F^{\cal L}_{i\overline{j}}= -{1\over 4c} g_{i\overline{j}}$, where $g_{i\overline{j}}$ is the 2-point function of ${\cal N}=2$ chiral primaries with scaling dimension $\Delta=2$. In the case of ${\cal N}=2$ theories, the Zamolodchikov metric $G_{i\overline{j}}$ can be expressed in terms of $g_{i\bar j}$ as  $G_{i\overline{j}}=192 g_{i\overline{j}}$. Consequently, in a 4d ${\cal N}=2$ theory we have
$$
F^{{\cal L},{\cal N}=2}_{i\overline{j}} =  -{1\over 768 c} G_{i\overline{j}}
~,
$$
which agrees with the $\NN=1$ result  \eqref{finalcurvature1}.

Two comments are in order here:

\begin{itemize}
\item[$a)$] Notice that the computation of $F_{i\overline{j}}^{{\cal L}, {\cal N}=2}$ in \cite{Papadodimas:2009eu} was done by considering the relative holonomy between a chiral superconformal primary and its {\it first} superconformal descendant, while in this paper we are using a {\it second} superconformal descendant. While we do not include the detailed computation in this paper, we have checked that the result \eqref{finalcurvature1} for ${\cal N}=1$ can be also derived by using the first descendant, as in \cite{Papadodimas:2009eu}. 

\item[$b)$] If we have an ${\cal N}=2$ SCFT we can also view it as an ${\cal N}=1$ SCFT, where we simply ignore the extra supercharges. In this approach and for computations like the one we presented above, we need to be careful about the existence of operators which are protected from the point of view of ${\cal N}=2$, but seem ``accidentally protected'' from the point of view of ${\cal N}=1$ (for example, the hidden supercharges, or some of the currents of the enhanced R-symmetry). In some cases these ``accidental'' operators can contribute to the integrals of the curvature and modify the final result\footnote{For example, this is relevant when writing a 2d ${\cal N}=(4,4)$ theory as an ${\cal N}=(2,2)$ theory, see for instance \cite{Gomis:2016sab}.}. However, in this particular case we have checked that it does not happen. So the computation of the curvature of ${\cal L}$ in a 4d  ${\cal N}=2$ (or ${\cal N}=4$ SCFT) can be done either in exactly the way we presented above by writing the theory in ${\cal N}=1$ language, or as in \cite{Papadodimas:2009eu} using the full structure of ${\cal N}=2$ SUSY.

\end{itemize}

 \subsection{Mixed components of the curvature of ${\cal L}$ in 3d ${\cal N}=2$}

A similar computation can be performed in 3d $\NN=2$ SCFTs. We start with the superconformal Ward identity \eqref{mainward3d}
\be
\label{mainward3de}
\begin{gathered}
\langle \overline{\cal O}_m(\infty) {\cal O}_i(x) \overline{\cal O}_j(y) {\cal O}_k(z) \rangle =
\Big[(x-z)^2\Box_x \Box_z + 6(x-z)\cdot(\partial_x \Box_z -\partial_z \Box_x)\\
-18(\partial_x\cdot\partial_z)+12(\Box_x+\Box_z)\Big]\langle \overline{\phi}_m(\infty)  \phi_i(x)  \overline{\phi}_j(y) \phi_k(z)\rangle
~.
\end{gathered}
\ee
Making the replacement $\partial_\mu^z \rightarrow -\partial_\mu^x-\partial_\mu^y$ and sending $z\rightarrow 0$ we find
\be
\label{mainward3de1}
\begin{gathered}
\langle \overline{\cal O}_m(\infty) {\cal O}_i(x) \overline{\cal O}_j(y) {\cal O}_k(0) \rangle =
\Big[x^2\Box_x (\partial_x + \partial_y)^2 + 6x\cdot (\partial_x (\partial_x+\partial_y)^2 +(\partial_x+\partial_y)
\Box_x)\\
+18(\partial_x\cdot(\partial_x+\partial_y))
+12[\Box_x+(\partial_x+\partial_y)^2]\Big]\langle \overline{\phi}_m(\infty)  \phi_i(x)  \overline{\phi}_j(y) \phi_k(z)\rangle
~.
\end{gathered} 
\ee
In this case, the relation between the 2-point function coefficients of the exactly marginal operators and the chiral primaries with scaling dimension $\Delta=2$ is
\be
G_{i\overline{j}}=12 g_{i\overline{j}}
~.
\ee
So, as before, we can write
\be
G^{\overline{m}l}\langle \overline{\cal O}_m(\infty) {\cal O}_i(x) \overline{\cal O}_j(y) {\cal O}_k(0) \rangle= W(x,y) + g^{\overline{m}l} \langle \overline{\phi}_m(\infty) {\cal O}_i(x) \overline{\cal O}_j(y) \phi_k(0) \rangle
~,
\ee
where
\be
\begin{gathered}
W(x,y) \equiv 
\Big[ x^2\Box_x (\partial_x + \partial_y)^2 + 6x\cdot(\partial_x (\partial_x+\partial_y)^2 +(\partial_x+\partial_y)
\Box_x) \\
+18(\partial_x\cdot(\partial_x+\partial_y))+12(\partial_x+\partial_y)^2  \Big]G^{\overline{m}l} \langle \overline{\phi}_m(\infty) \phi_i(x) \overline{\phi}_j(y) \phi_k(0) \rangle
~.
\end{gathered}
\ee
The latter can be recast as
\begin{equation}
\label{Wrecasta}
    \begin{gathered}
    W(x,y) = 
    \Big[ \Box_x x^2 \Box_x + \Box_x \Box_y x^2 + 2(\partial_x\cdot \partial_y) x^2 \Box_x
    +8 (\partial_x\cdot x) \Box_x +2\Box_y  (\partial_x\cdot x)\\
    +2 \Box_x(\partial_y\cdot x)  +12 (\partial_x \cdot \partial_y) ( \partial_x\cdot x)
    -10 (\partial_x \cdot \partial_y) \Big] G^{\overline{m}l} \langle \overline{\phi}_m(\infty) \phi_i(x) \overline{\phi}_j(y) \phi_k(0) \rangle
    ~.
    \end{gathered}
\end{equation}
Next we use the following conformal Ward identity for scalar primaries of $\Delta=2$ in a 3d CFT\footnote{Again, this identity can be proved easily by direct computation after expressing the 4-point function in terms of the conformally-invariant cross-ratios}.
\be
\Box_x \langle\overline{ \phi}_m(\infty)\phi_i(x)\overline{\phi}_j(y)\phi_k(0)\rangle = {1\over x^2 |y|} \Box_y \left(|y|^3\langle\overline{ \phi}_m(\infty)\phi_i(x)\overline{\phi}_j(y)\phi_k(0)\rangle \right)
\ee
to rewrite Eq.\ \eqref{Wrecasta} as
\begin{equation}
    \begin{gathered}
    W(x,y) = 
    \bigg\{ \Box_x \Box_y y^2 + 2 \Box_x \partial_k^y y^k + \Box_x \Box_y x^2 + 2(\partial_x\cdot \partial_y) x^2 \Box_x\\
    +8 (\partial_x\cdot x) \left(\Box_y {y^2 \over x^2} + 2 \partial_k^y {y^k \over x^2}\right) +2\Box_y  (\partial_x\cdot x)+2 \Box_x(\partial_y\cdot x)  +12 (\partial_x \cdot \partial_y) ( \partial_x\cdot x)\\
    -10 (\partial_x \cdot \partial_y) \bigg\} G^{\overline{m}l} \langle \overline{\phi}_m(\infty) \phi_i(x) \overline{\phi}_j(y) \phi_k(0) \rangle
    ~.
    \end{gathered}
\end{equation}

These manipulations have allowed us to express $W$ as a double total derivative with respect to both $x$ and $y$
\begin{equation}
    W = \partial_\mu^y \partial_\nu^x H^{\mu\nu}(x,y)~,
\end{equation}
where
\begin{equation}
    \begin{gathered}
H^{\mu\nu}(x,y) = 
\Big[ \partial_y^\mu \partial_x^\nu y^2 +2 y^\mu \partial_x^\nu + \partial_y^\mu \partial_x^\nu x^2 + 2 \eta^{\mu\nu} x^2 \Box_x
+8 x^\nu\left(\partial_y^\mu {y^2 \over x^2} + 2 {y^\mu \over x^2}  \right)\\
+2 \partial^\mu_y x^\nu +2 \partial^\nu_x x^\mu +  12 \eta^{\mu\nu} (\partial_x \cdot x) -10 \eta^{\mu\nu}\big) \Big] G^{\overline{m}l} \langle \overline{\phi}_m(\infty) \phi_i(x) \overline{\phi}_j(y) \phi_k(0) \rangle
~.
\end{gathered}
\end{equation}
Notice that
\be
W(y,x) = \partial_\mu^x \partial_\nu^y H^{\mu\nu}(y,x) = \partial_\mu^y \partial_\nu^x H^{\nu\mu}(y,x)~,
\ee
hence we need to compute the quantity
\be
\begin{gathered}
F^{\cal L}_{i\overline{j}}  \delta^{l}_k={1\over 2} {1\over (2\pi)^2}\int_{|x|\leq 1} d^3x  \int_{|y|\leq 1} d^3y (W(x,y) -W(y,x)) 
\\
={1\over 2} {1\over (2\pi)^2}\int_{|x|\leq 1} d^3x \int_{|y|\leq 1}d^3y\,\,\, \partial_\mu^y\partial_\nu^x( H^{\mu\nu}(x,y)-H^{\nu\mu}(y,x))~.
\end{gathered}
\ee
Doing the integration by parts (and checking that there are no finite contributions from the limits $x \to 0$ and $y\rightarrow 0$ for reasons similar to the ones outlined in the 4d $\NN=1$ case) we find
\be
\label{onemore4}
F^{\cal L}_{i\overline{j}}  \delta^{l}_k={1\over 2} {1\over (2\pi)^2} \lim_{r\rightarrow 1^-}\int_{|y|=1}d\Omega_y\,\,\int_{|x|=r} d\Omega_x\,\,|x||y| y_\mu x_\nu   ( H^{\mu\nu}(x,y)-H^{\nu\mu}(y,x))~.
\ee
In the limit $r\rightarrow 1^-$, the contribution to the integral from angular regions on the spheres $|x|=1,|y|=1$, which have finite distance, cancels out after antisymmetrization. Similar to the 4d $\NN=1$ computation, the only possible contributions come from the region $x\approx y$ on the spheres. These can be computed using the OPE between the operators at $x,y$.

In 3d, analytic closed form expressions of the conformal blocks are not readily available, but we can consider separately the contribution of a general operator $\OO$ (not necessarily conformal primary) of dimension $\Delta$ and spin $s$ in the channel $(i\overline{j})\rightarrow (k\overline{l})$. We denote this as 
\be
\langle \overline{\phi}_m(\infty) \,\phi_i(x)\,   \overline{\phi}_j(y)\, \phi_k (0)\rangle=...+ C_{i\overline{j}}^{\cal O} C_{{\cal O} k \overline{m}} {1\over |x-y|^{4-\Delta}}{1\over |y|^\Delta}P_s(\cos\theta)  +... ~.
\ee
where $P_s(\cos\theta)$ is given by an (appropriately normalized) Legendre polynomial of $\cos\theta \equiv {(x-y)\cdot y \over |x-y||y|}$ which arises by contracting symmetric-traceless polynomials of rank $s$ in $(x-y)$ with those in $y$ \cite{Hogervorst:2013sma}. For the second term in \eqref{onemore4} we have a similar expansion interchanging $x,y$. Hence, we can write the final result for the curvature as
\be
F ={1\over 2}
G^{\overline{m}l} \times \sum_{\Delta,s} C_{i\overline{j}}^{\cal O} C_{{\cal O} k \overline{m}} \,\, X_{\Delta,s}
~,
\ee
where we have defined the kinematic quantity
\be
\begin{gathered}
X_{\Delta,s} \equiv {1\over (2\pi)^2}\lim_{r\rightarrow 1^-} \int_{|x|=r} d\Omega_2^x \int_{|y|=1}d\Omega_2^y\, |y| |x|  \times\\
 \times
\Big\{\Big[(x\cdot \partial_x)(y\cdot \partial_y) y^2 +2y^2(x\cdot \partial_x)
+(x\cdot\partial_x)(y\cdot \partial_y) x^2 + 2 (x\cdot y) x^2 \Box_x  +
8 (y\cdot \partial_y) y^2 +16 y^2\\+2 x^2 (y\cdot\partial_y)+12 (x\cdot y)(x\cdot\partial_x +3)
+2 (x\cdot\partial_x)(x\cdot y)-10 (x\cdot y)\Big]{{1\over |x-y|^{4-\Delta}}{1\over |y|^\Delta}}P_s(\cos\theta)
\\
 -(x \leftrightarrow y)\Big\}
 ~.
\end{gathered}
\ee
In the range of conformal dimensions allowed by the unitarity bounds we find that the only potential contributions can arise from operators with
\begin{enumerate}
\item $\Delta=1$, $s=0$ and coefficient $X_{1,0}=-8$.

\item $\Delta=2$, $s=1$ and coefficient $X_{3,1}=48$.
\end{enumerate}

In 3d ${\cal N}=2$ theories, neutral scalar operators of $\Delta=1$ are superconformal primaries of short multiplets containing flavor currents, see for example \cite{Cordova:2016emh}. Using the fact that marginal operators have to be neutral under flavor symmetries and following a similar argument as the one we employed in 4d we conclude that such scalar operators can not appear in the OPE of $\phi_i$ and $\overline{\phi}_j$.

For the same reason, in the case of spin 1 operators only the R-current can contribute. Considering the limit $x\rightarrow y$ we have
\be
\langle\overline{\phi}_m(\infty)\phi_i(x)\overline{\phi}_j(y)\phi_k(0)\rangle =  {R_i R_k\over 16\pi^2 C_V} g_{i\overline{j}} g_{k\overline{m}} {(x-y)\cdot y \over |x-y|^4 |y|^6}+...~.
\ee
Consequently,
\be
C^J_{i\overline{j}}C_{Jk\overline{m}}  =  - {R_i R_k\over 16\pi^2 C_V} g_{i\overline{j}} g_{k\overline{m}}
~.
\ee
For the operators of interest we have again $R_i=R_k=2$.

As a result, we find that the curvature is
\be
F_{i\bar j} \delta_k^l = {1\over 2} G^{\overline{m}l}\times 48 \times \left(-{4\over 16 \pi^2 C_V} g_{i \overline{j}} g_{k\overline{m}}\right)
\ee
or by using the relation $G_{i\overline{j}} = 12 g_{i\overline{j}}$,
\be
F_{i\overline{j}} = -{1\over 24 \pi^2 C_V} G_{i\overline{j}}
~.
\ee
It is convenient to express the coefficient $C_V$ in terms of the stress tensor two-point function coefficient $C_T$. We will use conventions where
 \be
 \label{conventionst}
 \langle T_{\mu\nu}(x) T_{\kappa\lambda}(0)\rangle = C_T {1\over |x|^{6}}I_{\mu\nu,\kappa\lambda}(x)
 \ee
 with
 \be  I_{\mu\nu,\kappa\lambda}(x) = {1\over 2}(I_{\mu\kappa}(x)I_{\nu\lambda}(x)+I_{\mu\lambda}(x)I_{\nu\kappa}(x))-{1\over 3}\delta_{\mu\nu}\delta_{\kappa\lambda})\qquad,\qquad I_{\mu\nu}(x) = \delta_{\mu\nu}-2{x_\mu x_\nu \over x^2}
 ~.
\ee
From Ref.\ \cite{Barnes:2005bm}, we observe that $C_V = {1\over 6} C_T$. So, all in all we obtain
\be
F_{i\overline{j}} = -{1\over 4 C_T \pi^2}G_{i\overline{j}}
~,
\ee.

\subsection*{Acknowledgments}

We would like to thank S. Komatsu, C. Rosen and S. Zhiboedov for discussions and comments. We would like to especially thank Marco Baggio for discussions and collaboration on related topics.

\begin{appendix}

\section{SUSY Ward identities}
\label{appendixward}

For the convenience of the reader, in this Appendix we summarize useful superconformal Ward identities in 4d $\NN=1$ SCFTs. Very similar Ward identities exist in 3d $\NN=2$ SCFTs, but we will refrain from spelling out their detailed form. Moreover, in this paper we are always considering correlators in the superconformal vacuum of the theory, which obeys
\be
Q_a \vac = \overline{Q}_{\dot{a}}\vac= S^a \vac = \overline{S}^{\dot{a}} \vac= 0
~.\ee
For simplicity, we do not write explicitly the bra and ket in correlators and use the notation
\be
\langle {\cal O}_1(x_1)...{\cal O}_n(x_n)\rangle \equiv \langle 0|  {\cal O}_1(x_1)...{\cal O}_n(x_n)\vac
~.\ee
\subsection{Ward identities for supercharges}

We start with the case of bosonic operators. Since  $\langle 0|Q_a=0$, we have
\be
\langle Q_a {\cal O}_1(x_1)...{\cal O}_n(x_n)\rangle = 0
~.
\ee
We start moving $Q_a$ towards the right until it annihilates the ket. In the process, using
\be
Q_a {\cal O} = [Q_a,{\cal O}] + {\cal O}Q_a 
\ee
we pick up commutators to get
\be
\label{wardqpre}
\sum_{k=1}^n \langle {\cal O}_1(x_1)...[Q_a,{\cal O}_k(x_k)]...{\cal O}_n(x_n)\rangle = 0
~.\ee
Since $[Q_a,{\cal O}(x)] = [Q_a,{\cal O}](x)$, this can also be written as
\be
\label{wardq}
\sum_{k=1}^n \langle {\cal O}_1(x_1)...[Q_a,{\cal O}_k](x_k)...{\cal O}_n(x_n)\rangle = 0
~.
\ee

If some of the operators are fermionic, then it is more useful to form the anticommutator when passing $Q_a$ towards the right. We can do it using
\be
Q_a {\cal O} = \{Q_a, {\cal O}\} - {\cal O}Q_a
\ee
Then we get an identity like \eqref{wardq} but with additional minus signs depending on the pattern of fermionic operators.

An obviously similar Ward identity holds for the supercharge $\overline{Q}_{\dot{a}}$.

\subsection{Ward identity for superconformal charges}

Following a similar reasoning for the superconformal charges $S^a$ we obtain the equivalent of \eqref{wardqpre} as
\be
\label{wardspre}
\sum_{k=1}^n \langle {\cal O}_1(x_1)...[S^a,{\cal O}_k(x_k)]...{\cal O}_n(x_n)\rangle = 0~.
\ee
Notice, however, that unlike the case $[Q_a,{\cal O}(x)] = [Q_a,{\cal O}](x)$, now we have $[S^a,{\cal O}(x)] = [S^a,{\cal O}](x) + i x^{\dot{a}a} [\overline{Q}_a,{\cal O}](x)$. Hence, the Ward identity \eqref{wardqpre} can be written as
\be
\label{wards}
\sum_{k=1}^n \langle {\cal O}_1(x_1)...\left([S^a,{\cal O}_k]+i x_k^{\dot{a}a} [\overline{Q}_{\dot{a}},{\cal O}_k](x_k)\right)...{\cal O}_n(x_n)\rangle = 0~.
\ee
The equivalent identity for $\overline{S}^{\dot{a}}$ is
\be
\label{wardsbar}
\sum_{k=1}^n \langle {\cal O}_1(x_1)...\left([\overline{S}^a,{\cal O}_k]-i x_k^{\dot{a}a} [Q_a,{\cal O}_k](x_k)\right)...{\cal O}_n(x_n)\rangle = 0~.
\ee
Again, if some operators ${\cal O}_i$ are fermionic we can derive similar identities with anticommutators and some additional minus signs.

\subsection{Combining the two superconformal Ward identities}

It is often useful to consider linear combinations of the previous Ward identities. In particular, consider the following combination of fermionic generators, which depends on an arbirarily chosen spacetime point $z$ that we can select to our convenience
$$
\q^a(z) \equiv S^a -i z^{\dot{a}a} \overline{Q}_{\dot{a}}
~.
$$
This operator obeys 
$$
\q^a(z) \vac =0 \qquad,\qquad\langle 0| \q^a(z) = 0
~.
$$
Hence, we have
\be
\sum_{k=1}^n \langle {\cal O}_1(x_1)...[\q^a,{\cal O}_k(x)]...{\cal O}_n(x_n)\rangle = 0
\ee
or using \eqref{wardsbar}
\be
\label{wardcomboq}
\sum_{k=1}^n \langle {\cal O}_1(x_1)...\left([S^a,{\cal O}_k](x_k)+i (x_k-z)^{\dot{a}a} [\overline{Q}_{\dot{a}},{\cal O}_k](x_k)\right)...{\cal O}_n(x_n)\rangle = 0~.
\ee
A useful application of this identity, that we use several times in the main text, is to select $z=x_{k_0}$ where $x_{k_0}$ is the spacetime coordinate $x_{k_0}$ of one of the operators ${\cal O}_{k_0}$ in the correlator. Then, as we can see above, the term $[\overline{Q}_{\dot{a}},{\cal O}_{k_0}(x_{k_0})]$ does not contribute. If ${\cal O}_{k_0}(x_{k_0})$ also happens to be annihilated by $S^a$, for example if it is superconformal primary, then that particular operator in the correlator does not contribute at all to the Ward identity.

We also have a similar Ward identity if we start with 
\be
\q^{\dot{a}}(z) \equiv \overline{S}^{\dot{a}} + i z^{\dot{a}a}Q_a
~,
\ee
which gives
\be
\label{wardcomboqbar}
\sum_{k=1}^n \langle {\cal O}_1(x_1)...\left([\overline{S}^{\dot{a}},{\cal O}_k](x_k)-i (x_k-z)^{\dot{a}a} [Q_a,{\cal O}_k](x_k)\right)...{\cal O}_n(x_n)\rangle = 0
~.
\ee

\section{Proof of 4-point function Ward identities with marginal operators}
\label{4pointwardproof}

\subsection{4d ${\cal N}=1$}

Consider the 4-point function
$$
\langle {\cal O}_i(x) \overline{\cal O}_j(y) \overline{\cal O}_k(z) {\cal O}_l(w)\rangle
~.
$$
Using the superconformal Ward identities of App.\ \ref{appendixward}, this can be related to the 4-point function of chiral/antichiral primaries $\langle \phi_i(x) \overline{\phi}_j(y) \overline{\phi}_k(z) \phi_l(w)\rangle$. In this Appendix we sketch the derivation. 

First, we notice that the marginal operator
\beq
\label{onorm}
{\cal O}_i(x) = {1\over 4} \epsilon^{ab} \{Q_a,[Q_b,\phi_i]\}(x)
\eeq
can also be written as
\beq
\label{onorma}
{\cal O}_i(x) =- {1\over x^2}  {1\over 4} \epsilon_{\dot{a}\dot{b}} \{\overline{S}^{\dot{a}},[\overline{S}^{\dot{b}},\phi_i(x)]\}
~.
\eeq
In the last formula \eqref{onorma} it is important to distinguish between the commutators $[\overline{S}^{\dot a},\phi_i(x)]$ and $[\overline{S}^{\dot a},\phi_i](x)$. Consequently,
$$
\langle {\cal O}_i(x) \overline{\cal O}_j(y) \overline{\cal O}_k(z) {\cal O}_l(w)\rangle =-
{1\over x^2}  {1\over 4}  \epsilon_{\dot{a}\dot{b}}\langle \left(\{\overline{S}^{\dot{a}},[\overline{S}^{\dot{b}},\phi_i(x)]\}\right)\overline{\cal O}_j(y) \overline{\cal O}_k(z) {\cal O}_l(w)\rangle~
~.
$$

The successive application of the superconformal Ward identities yields
\bea
\label{4ward11}
\langle {\cal O}_i(x) \overline{\cal O}_j(y) \overline{\cal O}_k(z) {\cal O}_l(w)\rangle &=&
-{1\over x^2}  {1\over 4}  \epsilon_{\dot{a}\dot{b}}\langle \phi_i(x) \left(\{\overline{S}^{\dot{a}},[\overline{S}^{\dot{b}},\overline{\cal O}_j(y)]\}\right) \overline{\cal O}_k(z) {\cal O}_l(w)\rangle
\nonumber\\
&&-{1\over x^2}  {1\over 4}  \epsilon_{\dot{a}\dot{b}}\langle \phi_i(x)  \overline{\cal O}_j(y) \left(\{\overline{S}^{\dot{a}},[\overline{S}^{\dot{b}},\overline{\cal O}_k(z)]\}\right) {\cal O}_l(w)\rangle             
\\
&&-2 {1\over x^2}  {1\over 4}  \epsilon_{\dot{a}\dot{b}} 
\langle \phi_i(x) \left([ \overline{S}^{\dot{a}},\overline{\cal O}_j(y)] \right)\left([\overline{S}^{\dot{b}},\overline{\cal O}_k(z)]\right) {\cal O}_l(w)\rangle
~.\nonumber
\eea
For the first and second terms on the RHS of \eqref{4ward11} we use the following result, which can be derived by straightforward application of the superconformal algebra
\be
\label{ssO}
\epsilon_{\dot{a}\dot{b}}\left(\{\overline{S}^{\dot{a}},[\overline{S}^{\dot{b}},\overline{\cal O}_j(y)]\}\right)
=-4 \left[y^2 \Box_y +8(y\cdot\partial_y)+24\right]\overline{\phi}_j(y)
~.
\ee
For the last term in \eqref{4ward11} we consider the 4-point function
$$
 \epsilon_{\dot{a}\dot{b}}  
 \langle \phi_i(x) \left([ \overline{S}^{\dot{a}},\overline{\cal O}_j(y)] \right)\left([\overline{S}^{\dot{b}},\overline{\cal O}_k(z)]\right) {\cal O}_l(w)\rangle      
 ~.
$$
We are interested in the limit $x\rightarrow \infty$, which exhibits some simplifications. For instance, if we move the $Q$'s away from $w$, we can ignore any terms where the $Q$'s act on $x$ as those will be subleading as $x\rightarrow \infty$. Also notice that if two Q's act on either the operator at $y$ or $z$, it will be annihilated. Hence, we only need to consider terms of the form
$$
\epsilon_{\dot{a}\dot{b}} \epsilon^{cd}
\langle \phi_i(x) \left(\{Q_c,[ \overline{S}^{\dot{a}},\overline{\cal O}_j(y)]\} \right)\left(\{Q_d,[\overline{S}^{\dot{b}},\overline{\cal O}_k(z)]\}\right) \phi_l(w)\rangle
$$
Like \eqref{ssO} this correlation function can be simplified using the relations of the superconformal algebra.

Using these manipulations, we can write the 4-point function of interest as a sum of three terms
\beq
\langle {\cal O}_i(\infty) \overline{\cal O}_j(y) \overline{\cal O}_k(z) {\cal O}_l(w)\rangle  = {\tt I}_1 + {\tt I}_2 + {\tt I}_3
~.
\eeq
For the first term, Eq.\ \eqref{ssO} yields
\be
\begin{gathered}
{\tt I}_1=- {1\over 4}  \epsilon_{\dot{a}\dot{b}}\langle \phi_i(\infty) \left(\{\overline{S}^{\dot{a}},[\overline{S}^{\dot{b}},\overline{\cal O}_j(y)]\}\right) \overline{\cal O}_k(z) {\cal O}_l(w)\rangle
\\
= \left[y^2 \Box_y +8(y\cdot\partial_y)+24\right]\langle \phi_i(\infty) \overline{\phi}_j(y) \overline{\cal O}_k(z) {\cal O}_l(w)\rangle
~.
\end{gathered}
\ee
By moving the Q's away from ${\cal O}_l(w)$, we obtain
\beq
{\tt I}_1 = \left[y^2 \Box_y +8(y\cdot\partial_y)+24\right]\Box_z\langle \phi_i(\infty) \overline{\phi}_j(y) \overline{\phi}_k(z) \phi_l(w)\rangle 
~.
\eeq
Similarly, for the second term we find
\beq
{\tt I}_2 = \left[z^2 \Box_z +8(z\cdot\partial_z)+24\right] \Box_y \langle \phi_i(\infty) \overline{\phi}_j(y) \overline{\phi}_k(z) \phi_l(w)\rangle 
\eeq
and finally for the third term
\beq
{\tt I}_3= -\left[32 (\partial_y \cdot \partial_z)+8 (z\cdot\partial_y)\Box_z+8 (y\cdot \partial_z)\Box_y+2(y\cdot z)\Box_y \Box_z \right]\langle \phi_i(\infty)\overline{\phi}_j(y)\overline{\phi}_k(z)\phi_l(w)\rangle  
~.
\eeq
The desired Ward identity is
\be
\label{4pointwardidentity1}
\begin{gathered}
\langle {\cal O}_i(\infty) \overline{\cal O}_j(y) \overline{\cal O}_k(z) {\cal O}_l(w)\rangle =\bigg\{ (y-z)^2 \Box_y \Box_z + 8 (y-z)\cdot(\partial_y\Box_z-\partial_z \Box_y)\\
-32 (\partial_y \cdot\partial_z) + 24(\Box_z + \Box_y)\bigg\}\langle \phi_i(\infty)\overline{\phi}_j(y)\overline{\phi}_k(z)\phi_l(w)\rangle
~.
\end{gathered}
\ee

Notice that it is possible to recast this identity in various other forms either by: $i)$ Pulling the derivatives through the coordinate factors or $ii)$ Using ordinary conformal Ward identities for the correlator $\langle \phi_i(\infty)\overline{\phi}_j(y)\overline{\phi}_k(z)\phi_l(w)\rangle$.

We consider two applications of this identity in the main text: $i)$ to confirm K\"ahlerity and $ii)$ to compute the curvature of $\LL$. For the first one we need to take $w\rightarrow 0$. For the second one we take $z\rightarrow 0$ and replace the $z$-derivatives with $w,y$ derivatives.

We also notice that by pulling all the derivatives to the left on the RHS, the formula can be written as
\be
\label{4pointwardidentity2}
\begin{gathered}
\langle {\cal O}_i(\infty) \overline{\cal O}_j(y) \overline{\cal O}_k(z) {\cal O}_l(w)\rangle =\bigg\{\Box_y \Box_z(y-z)^2  + 4 \Box_z (\partial_y\cdot(y-z))\\
-4 \Box_y (\partial_z \cdot(y-z))-8(\partial_y\cdot\partial_z) \bigg\}\langle \phi_i(\infty)\overline{\phi}_j(y)\overline{\phi}_k(z)\phi_l(w)\rangle
~.
\end{gathered}
\ee

\subsection{3d ${\cal N}=2$}

An analogous Ward identity can be derived in 3d $\NN=2$ SCFTs. Since many of the intermediate steps are very similar to the corresponding ones in the 4d $\NN=1$ discussion, we will omit many of the pertinent details. In 3d $\NN=2$ SCFTs the marginal operators take the form
\be
\label{ward3daa}
{\cal O}_i =  {1\over 4}\epsilon^{ab} \{ Q_a,[Q_b,\phi_i]\}
~,
\ee
where again we introduced a specific convenient normalization factor and by convention $\epsilon^{12}=1$. In addition, there is a complex-conjugate version of $\OO_i$. Similar to the expression \eqref{onorma}, we can also recast the exactly marginal operator into the form
\be
\label{ward3dab}
{\cal O}_i = -{1\over x^2}  {1\over 4}\epsilon_{ab}\{\overline{S}^a,[\overline{S}^b,\phi_i]\}
~,
\ee
where $\epsilon_{12}=-1$. Then, by successive application of the superconformal Ward identities we arrive at the analog of Eq.\ \eqref{4ward11}. We consider each term that arises in this expression separately.

The first term of interest involves result 
\be
\label{ward3dac}
\epsilon_{ab} \{\overline{S}^a,[\overline{S}^b,\overline{\cal O}_j(y)]\}= -4(y^2\Box_y + 6(y\cdot\partial_y) +12)\,\overline{\phi}_j(y)
~.
\ee
which follows from straightforward application of the superconformal algebra. 
The second term that arises in the computation involves the 4-point function
\be
\label{ward3dai}
\begin{gathered}
4 \epsilon_{ab} \epsilon^{cd} \langle \phi_i(x)
\left(\{Q_c,[S^a, \overline{\cal O}_j(y)]\} \right)  \left(\{Q_d,[S^b,\overline{\cal O}_k]\}(z)\right)\phi_l(w)\rangle \\
=- 16 
\bigg[ 18(\partial_y\cdot \partial_z)+6(y\cdot \partial_z)\Box_y+6(z\cdot\partial_y)\Box_z+2(y\cdot z)\Box_y \Box_z \bigg] \langle \phi_i(x) \overline{\phi}_j(y)\overline{\phi}_k(z) \, \phi_l(w)\rangle
~.
\end{gathered}
\ee

Using these results, and taking the $x\rightarrow \infty$ limit, we finally obtain the result
\be
\label{ward3dal}
\begin{gathered}
\langle {\cal O}_k(\infty) \overline{\cal O}_j(y)\overline{\cal O}_k(z) {\cal O}_l(w)\rangle =
\bigg\{ (y-z)^2\Box_y \Box_z \\
+ 6(y-z)\cdot(\partial_y \Box_z -\partial_z \Box_y)-18(\partial_y\cdot\partial_z)+12(\Box_y+\Box_z)\bigg\} \langle \phi_i(\infty) \overline{\phi}_j(y) \overline{\phi}_k(z) \phi_l(w)\rangle
~.
\end{gathered}
\ee

\section{Curvature components and the K\"ahler structure}
\label{kahlercurv}

In this Appendix we derive the identity \eqref{riemann3}, 
\beq
\label{kcurvaa}
R_{ij\bar k \bar l} = 0
~,
\eeq
for the Riemann tensor on a 4d $\NN=1$ superconformal manifold. This identity is one of the consequences of the K\"ahler structure of the superconformal manifold. The corresponding identity in 3d $\NN=2$ superconformal manifolds can be derived in an exactly analogous manner and will not be discussed here explicitly.

We first use the superconformal Ward identity \eqref{4pointwardidentity2} to rewrite the 4-point function \eqref{kahlercor} as
\be
\begin{gathered}
\langle {\cal O}_l(\infty) \overline{\cal O}_i(x) \overline{\cal O}_j(y) {\cal O}_k(0)\rangle =
\bigg\{\Box_x \Box_y(x-y)^2  + 4 \Box_y (\partial_x \cdot(x-y)) \\
-4 \Box_x (\partial_y \cdot(x-y))-8 (\partial_x\cdot\partial_y) \bigg\}\langle \phi_l(\infty)\overline{\phi}_i(x)\overline{\phi}_j(y)\phi_k(0)\rangle   
~.
\end{gathered}
\ee
Consequently,
\beq
\label{kproofaa}
\langle {\cal O}_l(\infty) \overline{\cal O}_i(x) \overline{\cal O}_j(y) {\cal O}_k(0)\rangle 
 = \partial^y_\mu \partial^x_\nu G^{\mu\nu}(x,y)
\eeq
where
\be
\label{kahlerproofg}
G^{\mu\nu}(x,y) =
\left\{\partial^\mu_y \partial^\nu_x (x-y)^2  + 4 \partial^\mu_y (x-y)^\nu -4 \partial^\nu_x  \cdot(x-y)^\mu-8\eta^{\mu\nu} \right\}\langle \phi_l(\infty)\overline{\phi}_i(x)\overline{\phi}_j(y)\phi_k(0)\rangle 
~.
\ee
Since the 4-point function \eqref{kproofaa} is integrated over $x$, $y$ in the curvature formula \eqref{riemannf}, we can use the total-derivative form on the RHS of Eq.\ \eqref{kproofaa} to evaluate the integral by integrating by parts. We need to analyze carefully the potential boundary contributions as $x,y\rightarrow 0$. As explained below Eq.\ \eqref{curvdefinition} there are no contributions from the coincidence limit $x\rightarrow y$.

\subsection*{Behaviour near $y\rightarrow 0$}

First we consider the singularity structure as $y\rightarrow 0$. For that we need to examine the integral
\be
I(x,\epsilon)\equiv \int_{\epsilon<|y|\leq 1} d^4y\left(
\langle {\cal O}_l(\infty) \overline{\cal O}_i(x) \overline{\cal O}_j(y) {\cal O}_k(0)\rangle- (x\leftrightarrow y)\right)
\ee
and study the potential divergences as $\epsilon\rightarrow 0$. These can be analyzed by considering the OPE of the operators at $y$ and $0$. Notice that since there is no scalar operator of dimension 4 in this OPE (or else the operators would not be exactly marginal), we do not expect any logarithmic divergences. Thus, we have
\be
I(x,\epsilon) = \sum_a {c_a \over \epsilon^{4-\Delta_a}} +\widetilde{I}(x) + O(\cal \epsilon) 
~,
\ee
where possible divergent terms are coming from operators with $\Delta_a<4$ in the OPE between marginal operators. We want to isolate the finite piece $\widetilde{I}(x)$.

To do this we can rewrite the above integral in total-derivative form as
\be
I(x,\epsilon) = \int_{\epsilon<|y|\leq 1} d^4y\,\, \partial_\mu^y V^\mu(x,y)
~,
\ee
where
\be
\label{kahlerproofv}
V^\mu = \partial_\nu^x G^{\mu\nu}(x,y) -\partial_\nu^x G^{\nu\mu}(y,x)
~.
\ee
Using the divergence theorem we find
\be
\label{kahlerproofi}
I(x,\epsilon) = \int_{|y|=1} d\Omega_y |y|^2y_\mu V^\mu - \int_{|y|=\epsilon} d\Omega_y |y|^2y_\mu V^\mu 
\ee
For the computation of the curvature we are instructed to expand the integral above in powers of $\epsilon$ as $\epsilon\rightarrow 0$, throw away the divergent pieces and keep the finite piece (as discussed above there are no log divergences). Taking into account the angular integral at fixed $|y|=\epsilon$ the only possible finite contribution to $\widetilde{I}(x)$ can arise if we there is a term of the form
\be
\label{potsing}
V^\mu =  {\rm const}\times {y^\mu \over |y|^4}+...
\ee
Inspecting the relation of $V^\mu$ to the 4-point function $\langle \phi\overline{\phi}\, \overline{\phi} \phi \rangle$ via $G^{\mu\nu}$, we infer that the only operators capable of contributing a term of the form \eqref{potsing} are operators in the $\overline{\phi}_k(y) \phi_l(0)$ OPE with integer $\Delta$ and $\Delta \leq 4$. This means that their spin obeys $s\leq 2$. Considering the explicit functional form of contribution of possible operators of scaling dimension $\Delta$ and spin $s\leq 2$ in the $\overline{\phi}_k(y) \phi_l(0)$ OPE, we find that none of them contributes a term of the form \eqref{potsing}. Thus, we conclude that there is no finite contribution in the $y\to 0$ limit and we have
\beq
\tilde{I}(x)\equiv \lim_{reg, \epsilon\rightarrow 0} I(x,\epsilon) = 
\int_{|y|=1} d\Omega_y |y|^2y_\mu V^\mu
~.
\ee
 
\subsection*{Behaviour near $x\rightarrow 0$}

Next, we consider the integral over $x$. We have
\be
H(\epsilon,r) = \int_{\epsilon<|x|\leq r} d^4x\, \widetilde{I}(x)
~,
\ee
where we first want to send $\epsilon\rightarrow 0$, throw away the divergent terms, and afterwards send $r\rightarrow 1^-$.

Writing $V^\mu = \partial_\nu^x G^{\mu\nu} - (x\leftrightarrow y)$ we obtain
\be
\begin{gathered}
H(\epsilon,r) = \int_{|y|=1}d\Omega_y|y|^2 y_\mu \times\\
\times\left\{\left[\int_{|x|=r}d\Omega_x |x|^2 x_\nu (G^{\mu\nu}(x,y) - (x\leftrightarrow y))\right]-\left[\int_{|x|=\epsilon}d\Omega_x |x|^2 x_\nu (G^{\mu\nu}(x,y) - (x\leftrightarrow y))\right] \right\} 
~.
\end{gathered}
\ee
There are finite contributions in the limit $\epsilon \rightarrow 0$, if the integrand has a term of the form 
\be
{x^\nu \over |x|^4}
~.
\ee
Examining the potential contribution of operators with $\Delta\leq 4$ and spin $l\leq 2$, we find that none of them can arise. As a result, there are no finite contributions from the limits $x,y\rightarrow 0$. 

\subsection*{Boundary terms from $|x|,|y|\rightarrow 1$}

The above considerations have reduced the computation of the curvature components of interest to a double integral over a boundary term
\be
R = \lim_{r\rightarrow 1^-} \int_{|x|=r}d\Omega_x
\int_{|y|=1} \Omega_y |x|^2 |y|^2 x^{\mu}y^{\nu} (G^{\mu\nu}(x,y) - G^{\nu\mu}(y,x))
\ee
with explicit form
\be
\begin{gathered}
R = 
\lim_{r\rightarrow 1^-} \int_{|x|=r}d\Omega_x
\int_{|y|=1} \Omega_y |x|^2 |y|^2
\bigg[ (x\cdot\partial_x)(y\cdot\partial_y) (x-y)^2\\
+4 (y\cdot\partial_y) (x^2-x\cdot y)-4 (x\cdot \partial_x)(x\cdot y -y^2) -8 x\cdot y) \bigg] \langle \phi_i(\infty)\overline{\phi}_j(x)\overline{\phi}_k(y)\phi_l(0)\rangle  
-(x\leftrightarrow y)
~.
\end{gathered}
\ee
In the limit $r\rightarrow 1^-$, the contribution to the above integrals of any region where $\Omega_x\neq \Omega_y$ vanishes, due to the antisymmetrization. The only possible contribution comes from regions $\Omega_x$ near $\Omega_y$, where the $r\rightarrow 1^-$ limit can not be taken inside the integral. For that region we can use the OPE in the channel $(\overline{j} \,\overline{k})\rightarrow (i\, l)$. Inspecting the form of the integrand, we notice that only terms with a singularity at least as strong as ${1\over |x-y|^3}$ have a chance of contributing. This means that such terms must originate from operators with $\Delta \leq 3$ in the OPE $\overline{\phi}_j \overline{\phi}_k$. Since both of these operators are antichiral primaries, the lowest scaling dimension that can appear on the RHS is $\Delta=6$. Consequently, the double integral is zero in the limit $r\rightarrow 1^-$.

In summary, we have verified that the curvature components in Eq.\ \eqref{kcurvaa} are indeed zero in accordance with the arguments about the K\"ahler structure in section \ref{kaehlerproof}.

\section{On the additivity of holonomies}
\label{appadd}

In subsection \ref{kaehlerproof} we argued that the holonomy of the holomorphic part of the tangent bundle splits into the separate holonomies of the line bundle $\LL^2$ and the vector bundle $\VV$. A similar statement applies to the anti-holomorphic part. This requires that the connection of the exactly marginal operators $\OO_i= \frac{1}{4} \epsilon^{ab} \{ Q_a,[Q_b,\phi_i]\}$ is the sum of the connection for the chiral primaries $\phi_i$ and the connection for the two $Q$'s. A sufficient condition for this is that the covariant derivative of 3-point functions of the form $\langle \overline{\OO}(x) G_{\mu,a}(y) [Q_b,\phi](z)$ must vanish. In this appendix we elaborate on this requirement.
 
As a warmup, let us first consider the additivity of holonomies for the {\it first} superconformal descendant. For that we consider the 3-point function
$$
\langle [\overline{Q}_{\dot{a}}, \overline{\phi}](x) \, G_{\mu,a}(y) \,\phi(z)\rangle
~,
$$
where the supercurrent is $G_{\mu,a} = [Q_a, J_\mu]$ with $J_\mu$ the $U(1)$ R-current. We want to prove that this 3-point function is covariantly constant. We will show this by inserting a marginal operator and proving that the resulting 4-point function is identically zero. 

\subsection*{Anti-Holomorphic deformation}

Consider the 4-point function
\be
\label{okokok}
{\rm I}=\langle \overline{\cal O}(w)\,\,[\overline{Q}_{\dot{a}}, \overline{\phi}](x) \, G_{\mu,a}(y) \,\phi(z)\rangle
~.
\ee
We select to first move a $\overline{Q}$ away from $\overline{\cal O}$, whose spinor index is different from the $\dot{a}$ of the operator at $x$.  We do this by selecting a linear combination of superconformal Ward identities so that $\overline{Q}$ does not act on $y$. The operator $\phi(z)$ is annihilated by both $\overline{Q},S$ so we do not get any contribution from it.
We will then get two types of terms, one from $S$ acting on $G_{\mu,a}(y) = [Q_a,J_\mu(y)]$, that gives back $J_\mu$
\be
{\rm I}_1=\langle [\overline{Q}_{\dot{a}},\overline{\phi}(w)]\,\,[\overline{Q}_{\dot{a}}, \overline{\phi}](x) \, J_{\mu}(y) \,\phi(z)\rangle
~,
\ee
and another term from $\overline{Q}$ acting on the operator at $x$, which will look like
\be
{\rm I}_2 =\langle [\overline{Q}_{\dot{a}},\overline{\phi}(w)]\,\,\{\overline{Q}_{\dot{b}},[\overline{Q}_{\dot{a}}, \overline{\phi}]\}(x) \, G_{\mu,a}(y) \,\phi(z)\rangle~.
\ee
To show that ${\rm I}_1$ vanishes, we move the  $\overline{Q}_{\dot{a}}$ away from $w$, using a superconformal Ward identity so that we get no contribution from $J_\mu(y)$, which is annihilated by $S$. Again there is no contribution from $\phi(z)$ and the  contribution from $x$ also vanishes, since we have two supercharges $\overline{Q}_{\dot{a}}$ with the same spinor index acting on the operator.

To show that ${\rm I_2}$ vanishes, 
we move $\overline{Q}_{\dot{a}}$ away from $w$, arranging that $\overline{Q}$ does not act on $y$ (so only $S$ will act on $y$). We do not get a contribution from $x$ since there is already a $\overline{Q}_{\dot{a}}$ there. Hence, we get that ${\rm I}_2$ is proportional to the correlator ${\rm I}_2'$ given by
\be
{\rm I}_2' = \langle \overline{\phi}(w)\,\,\{\overline{Q}_{\dot{b}},[\overline{Q}_{\dot{a}}, \overline{\phi}]\}(x) \, J_{\mu}(y) \,\phi(z)\rangle~.
\ee
Now we move each of the two $\overline{Q}$'s from $x$, with a Ward identity so that $\overline{Q}$ does not act on $w$, and then both $\overline{Q}$'s will end on $J_\mu$. Since $J_\mu$ is a short multiplet, when two $\overline{Q}$'s act on it we get zero. Hence ${\rm I_2}'$ (and thus ${\rm I}_2$) vanish. So all in all the original correlator \eqref{okokok} vanishes.

\subsection*{Holomorphic deformation} 

Similarly for the holomorphic deformation consider the 4-point function
\beq
\mathbb{I}=\langle{\cal O}(w)\,\,[\overline{Q}_{\dot{a}}, \overline{\phi}](x) \, G_{\mu,a}(y) \,\phi(z)\rangle
~.
\ee
First we move $Q$ away from ${\cal O}$. We select a linear combination of Ward identities so that we do not get any contribution from $\phi(z)$. The operator
$G_{\mu,a}(y)=[Q_a,J_\mu(y)]$ is annihilated by $Q$, because it is a semi-short multiplet, and by $\overline{S}$, because $J_\mu$ is superconformal primary. As a result, we obtain a sole contribution from the term at $x$. Both $Q$ and $\overline{S}$ remove the $\overline{Q}$, hence the computation is reduced to the evaluation of the 4-point function
\be
\mathbb{I}' = \langle [Q_c,\phi](w) \overline{\phi}(x) G_{\mu,a}(y) \phi(z)\rangle
~.
\ee
Repeating the same analysis by moving $Q_c$ away from $w$ we finally obtain $\mathbb{I}'=0$.

\vspace{0.5cm}

Similar arguments can be applied to the second descendants that appear in the exactly marginal deformations. More specifically, we can similarly show that the 3-point function
\be
\langle \{\overline{Q}_{\dot{a}},[\overline{Q}_a,\overline{\phi}]\}(x)             G_{\mu,a}(y)  [Q_b,\phi](z)\rangle
\ee
is covariantly constant.

\subsection*{Holomorphic deformation}

Consider the 4-point function
\be
\langle{\cal O}(w) \{\overline{Q}_{\dot{a}},[\overline{Q}_a,\overline{\phi}]\}(x)             G_{\mu,a}(y)  [Q_b,\phi](z)\rangle
~.
\ee
First, we move $Q$ away from ${\cal O}$. It acts as a linear combination of $Q,\overline{S}$:
\begin{enumerate}
 \item We select the linear combination of Ward identities so that we do not get any contribution from the operator at $z$.
 \item As before, $G_{\mu,a}(y)$ is annihilated by both $Q$ and $\overline{S}$. 
 \item We obtain a single contribution from the term at $x$. Both $Q$ and $\overline{S}$ remove one $\overline{Q}$. 
 \item When we apply the above steps for both $Q$'s in ${\cal O}$ we end up with a correlator of the form
\be
\langle \phi(w) \overline{\phi}(x) G_{\mu,a}(y) [Q_a,\phi](z)\rangle
~.
\ee
In this 4-point function we move $Q_a$ away from $z$ using a Ward identity so that $Q$ does not act at $w$. For the reasons mentioned previously the correlator vanishes.
\end{enumerate}

\subsection*{Anti-Holomorphic deformation}

Finally, we consider the 4-point function
\be
\langle \overline{\cal O}(w) \{\overline{Q}_{\dot{a}},[\overline{Q}_{\dot{b}},\overline{\phi}]\}(x)             G_{\mu,a}(y)  [Q_b,\phi](z)\rangle
~.
\ee
We start by moving $\overline{Q}$ away from $\overline{\cal O}$.  It acts as a linear combination of $\overline{Q},S$ and gives no contributions from the operator at $x$. We select the Ward identity so that $\overline{Q}$ does not act at $y$.

We then obtain two types of correlators
\be
\langle [\overline{Q}_{\dot{c}},\overline{\phi}(w)] \{\overline{Q}_{\dot{a}},[\overline{Q}_{\dot{b}},\overline{\phi}]\}(x)             J_{\mu}(y)  [Q_b,\phi](z)\rangle
\ee
and
\be
\langle [\overline{Q}_{\dot{c}},\overline{\phi}(w)] \{\overline{Q}_{\dot{a}},[\overline{Q}_{\dot{b}},\overline{\phi}]\}(x)             G_{\mu,a}(y)  \phi(z)\rangle
~.
\ee
Moving the $\overline{Q}_{\dot{c}}$ away from $w$ we obtain correlators of the form
\be
\langle \overline{\phi}(w)  \{\overline{Q}_{\dot{a}},[\overline{Q}_{\dot{b}},\overline{\phi}]\}(x)             J_{\mu}(y)  \phi(z)\rangle
~.
\ee
Next we move $\overline{Q}_{\dot{a}}$ away from $x$. We arrange the Ward identity so that we $\overline{Q}$  does not act on the operator at $w$. That yields a correlation function of the form
\be
\langle \overline{\phi}(w) [\overline{Q}_{\dot{b}},\overline{\phi}](x) [\overline{Q}_{\dot{a}},J_\mu](y) \phi(z)]\rangle
~.
\ee
Repeating the process for $\overline{Q}_{\dot{b}}$ we end up with the 4-point function
\be
\langle \overline{\phi}(w) \overline{\phi}(x) \{\overline{Q}_{\dot{b}},[\overline{Q}_{\dot{a}},J_\mu]\}(y) \phi(z)]\rangle
~,
\ee
which vanishes because of shortening conditions of $J$.

\end{appendix}

\bibliography{khodge}
\bibliographystyle{utphys}

\end{document}